%% file: novel-fermi-function.tex
\documentclass[sn-mathphys,Numbered]{sn-jnl}
\usepackage{graphicx}%
\usepackage{multirow}%
\usepackage{amsmath,amssymb,amsfonts}%
\usepackage{amsthm}%
\usepackage{mathrsfs}%
\usepackage[title]{appendix}%
\usepackage{xcolor}%
\usepackage{textcomp}%
\usepackage{manyfoot}%
\usepackage{booktabs}%
\usepackage{algorithm}%
\usepackage{algorithmicx}%
\usepackage{algpseudocode}%
\usepackage{listings}%
\usepackage[normalem]{ulem}
\usepackage{comment}
\usepackage{subcaption}
\usepackage{wrapfig}
\newcommand{\orcidicon}{\includegraphics[width=0.32cm]{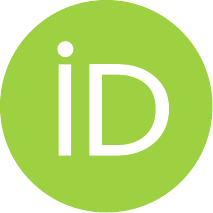}}
\newcommand{\orc}[1]{\href{https://orcid.org/#1}{\orcidicon}}

\newcommand{\orcA}{0000-0001-8217-1484}
\newcommand{\orcB}{0000-0001-5038-8427}
\newcommand{\orcC}{0000-0001-5474-2649}
\newcommand{\orcD}{0000-0003-2704-6474}
\newcommand{\orcE}{0000-0002-2289-4856}

\newcommand{\req}[1]{Eq.\,(\ref{#1})}
\newcommand{\rf}[1]{Figure~{\ref{#1}}}

\newcommand*{\MeV}{\text{ MeV}}

\DeclareMathOperator{\sgn}{sgn}

\newcommand*{\xred}{\color{black}}

\newtheorem{theorem}{Theorem}
\newcommand{\rTh}[1]{Theorem~{\ref{#1}}}

\newtheorem{lemma}[theorem]{Lemma} 

\newtheorem{remark}{Remark}

\begin{document}

\title[Article Title]{\xred Fermi-Dirac Integrals in Degenerate Regimes:\\ A Novel Asymptotic Expansion}
  
\author[1]{\fnm{Jeremiah} \sur{Birrell\orc{\orcE}}}
\author[2,3]{\fnm{Martin} \sur{Formanek\orc{\orcD}}}
\author[2]{\fnm{Andrew} \sur{Steinmetz\orc{\orcC}}} 
\author[2]{\fnm{Cheng Tao} \sur{Yang\orc{\orcB}}}
\author[2]{\fnm{Johann} \sur{Rafelski\orc{\orcA}}}


\affil[1]{\orgdiv{Department of Mathematics}, \orgname{Texas State University}, \city{\newline San Marcos}, \state{Texas}, \postcode{78666}, \country{USA}}

\affil[2]{\orgdiv{Department of Physics}, \orgname{The University of Arizona}, \city{\newline Tucson}, \state{Arizona}, \postcode{85721}, \country{USA}}

\affil[3]{\orgdiv{ELI Beamlines Facility}, \orgname{The Extreme Light Infrastructure ERIC}, \orgaddress{ \postcode{\newline 252 41} \city{Doln\'{i} B\v{r}e\v{z}any}, \country{Czech Republic\vskip 1.6cm}}}

\abstract{We characterize in a novel manner the physical properties of the low temperature Fermi gas in the degenerate domain as a function of temperature and chemical potential. For the first time we obtain low temperature $T$ results in the domain where several fermions are found within a de Broglie spatial cell. In this regime, the usual high degeneracy Sommerfeld expansion fails. The other known semi-classical Boltzmann domain applies when fewer than one particle is found in the de Broglie 
cell. {\xred We also improve on the understanding of the Sommerfeld expansion in the regime where the chemical potential is close to the mass and also in the high temperature regime. In these calculcations we use a novel characterization of the Fermi distribution
allowing the separation of the finite and zero temperature phenomena.} The relative errors of the three approximate methods (Boltzmann limit, Sommerfeld expansion, and the new domain of several particles in the de Broglie cell) are quantified.
}
\date{May 2024, To be published in the International Journal of Theoretical Physics}


\maketitle
\vskip -11.2cm 
{\xred 
\centerline{May 2024} 
\centerline{To be published in the {\it International Journal of Theoretical Physics}}
\vskip 0.4cm
\centerline{\bf Dedication:}
\vskip 0.1cm
\centerline{\bf We dedicate this work to  Louis de Broglie.}
\centerline{\it 100 years ago Louis de Broglie recognized the wave character of matter,} \centerline{\it the linchpin in the physical interpretation of our results.} 

}
\vskip 8.1cm
\section{Introduction}\label{Intro}
\subsection{Motivation}\label{Motivation}
Many interesting effects related to the Pauli principle appear at finite, but low temperatures, in vastly different physical systems~\cite{bludman1977equation,Elze:1980er,Ferrer:2019xlr,Kuebler:2021Th}. The limit $T\to 0$ of the Fermi-Dirac (FD) quantum system is subtle. We show analytical results which depend on particle number density within the de Broglie volume $V_\mathrm{dB}$, in $D$-dimensions defined by
\begin{equation}\label{dBVol}
V_\mathrm{dB}=L_\mathrm{dB}^D, \qquad L_\mathrm{dB}=
\frac{2\pi\hbar}{\langle p\rangle_T},\qquad \langle p\rangle_T=
\left\langle \sqrt{E^2-m^2}\ \right\rangle_T \;,
\end{equation}
 where angle brackets denote thermal averages. Any physical systems maintaining a prescribed particle number density in the $T\to 0$ limit requires a simultaneous chemical potential $\mu$ limit. For this reason, key physical features of low-temperature FD quantum gasses can be missed if the appropriate relation between $\mu$ and $T$ is not maintained in the low temperature limit. In this work, we show that a careful treatment of the simultaneous limit leads to novel asymptotic expansions of FD thermal averages. While in principle this is not an unexpected finding, we have found no mention of such mathematical considerations in literature.

The asymptotic limit of integrals of the FD distribution~\cite{dingle1957fermi,dingle1973asymptotic} are of interest to physicists studying many diverse statistical quantum systems~\cite{10.1063/1.1350634,Lourenco:2007zza,10.1063/1.1665160,FUKUSHIMA2014417,GIL2022126618,GIL2023108563} and behavior around the Fermi surface~\cite{kim2008notes,PhysRevB.103.205154}. Other use cases include compact astrophysical systems (white dwarfs, neutron stars, quark stars)~\cite{Kaspi:2017fwg,Ferrer:2019xlr,Ferrer:2023pgq}, Early Universe phenomenology~\cite{Rafelski:2021aey,Rafelski:2023emw,Grayson:2023flr,Steinmetz:2023nsc}, quark-gluon plasmas (QGP)~\cite{Elze:1980er,Letessier:2002ony,Rafelski:2020ajx,Yang:2021bko}, and in any situation when finite temperature Fermi effects are important. When investigating the low temperature limit of a Fermi gas, a systematic approach to approximating the physical integral involves employs the Sommerfeld expansion. It is worth noting that the Sommerfeld expansion, as described in Ref.~\cite{landau2013statistical}, is asymptotic in nature. It is within this context that we often see matter-of-fact statements that the ``Sommerfeld expansion fails'' in various regimes. A textbook example is seen on page 38 of Ref.~\cite{Kuebler:2021Th} (for the Sommerfeld expansion in this text, see p.~24). However, there appears to have been no progress towards a wholly distinct expansion which could rectify this issue. This situation seems to have lasted now for nearly a century. 

After several in-depth reviews of literature, we conclude that a full understanding of the domain of validity of the Sommerfeld expansion in the $(T,\Delta\mu)$-plane
\begin{equation}\label{DeltaMu}
\Delta\mu\equiv \mu-m\,,
\end{equation}
is lacking. The Sommerfeld expansion fails for regimes where the corrections to the $T\rightarrow0$ limit cannot be taken to be small. This makes it difficult to search for alternative expansions that are valid in other regions of the $(T,\Delta\mu)$-plane when dealing with multiple scales. One generally requires specialized asymptotic expansions that are adapted to specific regimes under consideration. This is particularly relevant when $\Delta\mu$ is small, a regime that the traditional Sommerfeld expansion was not designed for and which we consider here. Specifically, expanding the Sommerfeld result around the small chemical potential may lead to encountering singularities which underscore the necessity for a better tool to study this scenario. 

In this work we explore the different domains of low temperature FD gas, characterize the domain of validity of Sommerfeld expansion in physical terms, and obtain a novel approximation applicable to domains of statistical parameters where the ``expansion fails''. As a side-line result, we also provide a novel form of FD distribution that manifestly separates the zero and finite temperature contributions so that the finite temperature correction can be studied on its own in more complex environments than considered in our current method-oriented presentation. 

Our own interest and motivations to explore FD systems originated in cosmological computations involving both high and low temperature physics, such as comic neutrino background~\cite{Birrell:2013gpa,Birrell:2012gg} and cosmic magnetism~\cite{Steinmetz:2023nsc}. In the context of magnetism an added difficulty is the Landau discrete sum over Landau orbitals. In the regime of neutrino cosmology, considering their likely mass, the present-day free-streaming neutrinos exhibit characteristics of being cold and nonrelativistic fermions, with a small chemical potential reflecting the baryon asymmetry in the universe. During the Universe expansion neutrino abundance per quantum correlation de Broglie volume decreases {\xred when the ambient temperature scale and neutrino mass scale coincide}. Therefore the free-streaming neutrinos might fall into the novel regime described here {\xred for quite a long period of Universe expansion. We note that a related technique to ours, for the case of number density applied to study a degenerate gas in a freely expanding universe, can be found in \cite{Trautner:2016ias} in the limit in which free-streaming phenomena do not matter. 

Another natural area of physical relevance involves `warm' nuclear matter (temperature well below transformation to quark-gluon plasma). An estimation of the number of indistinguishable nucleons in the de Broglie volume at a normal nuclear matter density produces a value $\simeq 5$, consistent with the general understanding of atomic nuclei as being quantum objects, rather than assembly of classical interacting balls. This means that neutron stars in their formation history, and during neutron star mergers evolve through the warm nuclear matter non-Sommerfeld and non-Boltzmann regime we characterize. For in depth discussion of the current status of this field see for example Refs.\,\cite{Oertel:2016bki,Perego:2019adq,MUSES:2023hyz} and references therein. We further note recent interest in the understanding of properties of warm nuclear matter~\cite{Mroczek:2024sfp}, which is at present limited to the Sommerfeld expansion. Our newly identified regime and associated asymptotic expansion present an opportunity to further refine this work.
 
The newly identified regime is transient, however, it could influence the dynamical evolution or radial dependence of dense nuclear systems  which impact the physical outcome concerning the formation of neutron stars, their size and mass, or the detailed questions concerning dynamical observations involving neutron star mergers~\cite{Raithel:2019gws,Raithel:2021hye,Blacker:2023afl,Raithel:2023zml}. Similarly, the results we present could influence the interpretation of laboratory experiments involving the study of dense warm nuclear matter in relativistic heavy ion collisions, see Ref.\,\cite{Sorensen:2023zkk}, and references therein. 
 
Our method can be applied to general observables in these diverse cases. We intend to use the techniques developed here in future application-oriented work. This work could address applications in cosmology, in the study of the case of free-streaming neutrinos, which is of profound relevance for better understanding the evolution of the Universe. Furthermore our techniques are applicable to dense Fermi systems in {\it e.g.\/} stellar environments, and to dense Bose systems.}

\subsection{Domains of interest and their characteristic particle density}\label{domains}
In this work we derive and contrast expansions of FD matter in the following regimes which, to the best of the authors' knowledge, contain several novel insights. In \rf{fig:Thm3_vs_Sommerfeld_regions} we show the domains of interest in the ($T,\Delta\mu$)-plane (in units of $m$). The darker background shows regions with relative error less than $10\%$, lighter background shows error less $20\%$, when computing $\langle G\rangle_T$ for the illustrative choice $G(p)=1$. {\xred We emphasize that other choices of $G$ will lead to different domains; see Section \ref{ssec:NoChem} for further examples.} 

\begin{figure}[bht] 
\centering
 \includegraphics[width=.7\textwidth]{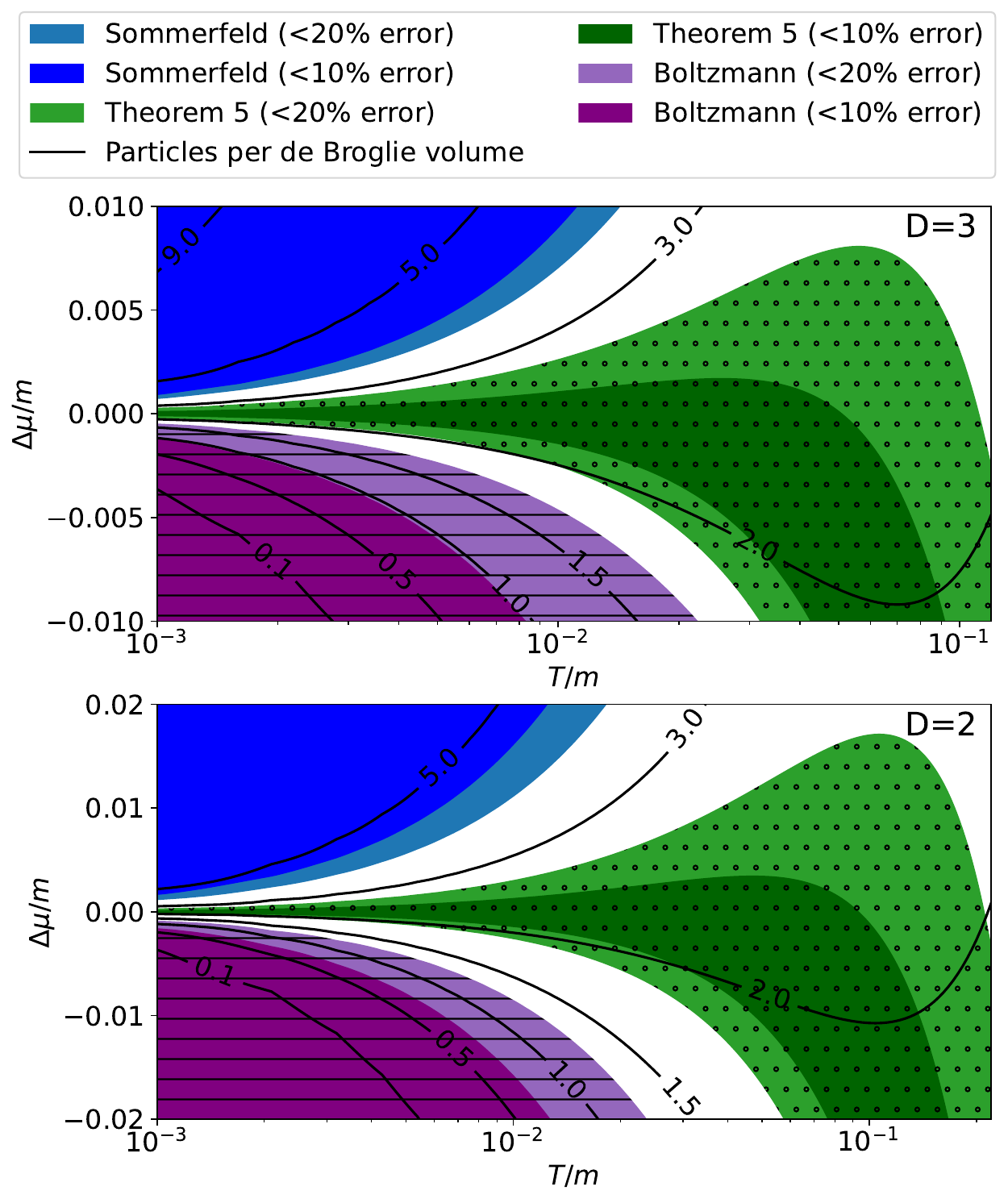}
\caption{Domains of interest spanned in the $(T/m,\Delta\mu/m\equiv(\mu-m)/m)$-plane, top for three dimensional and bottom frame for two dimensional systems: Sommerfeld expansion (blue), Boltzmann approximation (lined purple), and the new \rTh{thm:mu_zero_faster} domain (dotted green). Solid black contours show the number of particles per de Broglie volume \req{dBVol}. See text for a more detailed discussion.}\label{fig:Thm3_vs_Sommerfeld_regions}
\end{figure}

Errors here, and elsewhere in this work, were obtained by comparing the approximations with the result obtained from standard numerical quadrature; given the form of the FD integrand, such methods can be reliably used as long as the values of $\mu$ and $T$ are not too extreme. We also show using solid black contours lines in \rf{fig:Thm3_vs_Sommerfeld_regions} the number of particles per de Broglie volume. From these one can recognize
\begin{enumerate}
\item
The traditional Sommerfeld asymptotic expansion applies when statistical parameters allow for a quantum degenerate density of Fermi particles, with about 5 or more particles found in the corresponding de Broglie volume. As will be considered here the validity of this regime can be extended to the dense relativistic quantum domain. 
\item 
The domain of validity of the newly characterized dilute quantum region (\rTh{thm:mu_zero_faster}) is found when the combination of statistical parameters allows a moderate particle density of a few (2 to 3) particles per de Broglie volume, where quantum effects are expected to be nontrivial but different from properties of dense FD matter. 
\item
The Boltzmann limit applies to systems with about 1 or fewer particles per de Broglie volume. 
\end{enumerate}
Particle abundance described and shown in \rf{fig:Thm3_vs_Sommerfeld_regions} does not account for spin and other degeneracies. Finally we note that \rf{fig:Thm3_vs_Sommerfeld_regions} focuses on the non-relativistic domain; for a similar figure that shows the relativistic regime see \rf{fig:Thm3_vs_Sommerfeld_regions_terms_comp_deBroglie}.

\subsection{\xred Specific examples independent of chemical potential}\label{ssec:NoChem}
{\xred
When analyzing physical properties of a quantum gas, it is often of interest to express all results in terms of temperature, and particle number density rather than the grand canonical chemical potential. This approach aligns more closely with experimental conditions, as temperature and particle number density are directly measurable quantities, and particle density is often created by physical circumstance. In this section, we present results that are independent of the chemical potential providing a more practical understanding of the Fermi gas in different physical regimes.

In \rf{fig:2-NTT} we present the domains of validity as defined by the number density on the plane parameterized by $T/m$ and particle number per de Broglie volume for $D=2$ (bottom frame) and $D=3$ (top frame) dimensional systems. The black solid contours indicate the kinetic energy per particle in units of $T$. For $D=3$ we see that the traditional Sommerfeld asymptotic expansion is applicable for $T\ll m$, when particle number per de Brogile volume is greater than $4.5$, which is for example the case for normal and gravitationally compressed nuclear matter. 

\begin{figure} 
\centering
 \includegraphics[width=.8\textwidth]{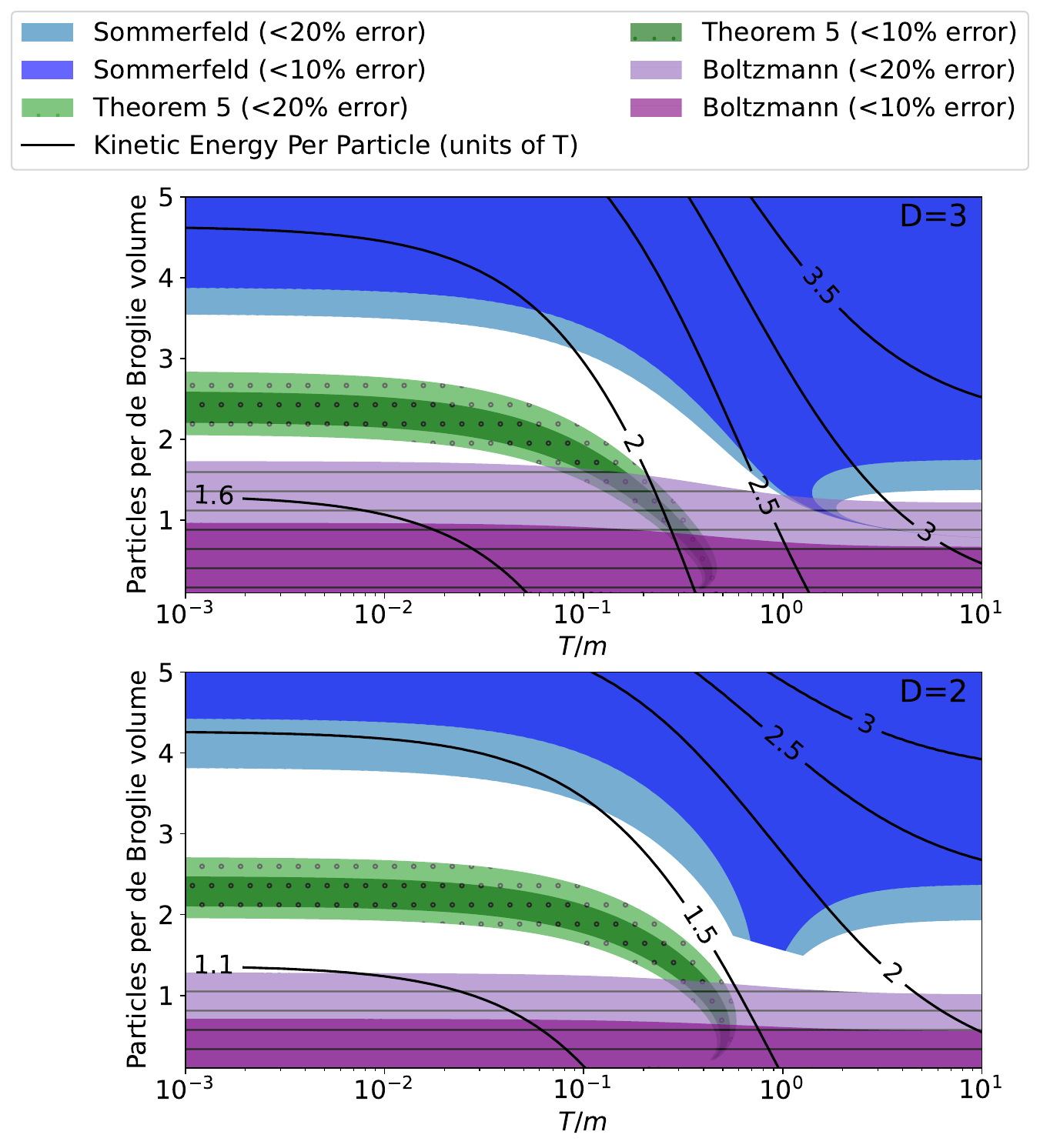}
\caption{{\xred
Study of energy per particle regimes of an ideal Fermi gas in the plane spanned by the number of particles in de Broglie volume, and temperature in units of $m$, $T/m$; top frame for the three dimensional $D=3$, and bottom frame for the two dimensional $D=2$ systems: Sommerfeld expansion (blue), Boltzmann approximation (lined purple), and the new \rTh{thm:mu_zero_faster} domain (dotted green). Solid black contours show the kinetic energy per particle in units of $T$. The darker color applies to domains with less than 10\% error, while the lighter color allows up to 20\% error.}}\label{fig:2-NTT}
\end{figure}

For the case $T\gg m$ the domain of validity of the Sommerfeld regime decreases to be as low as $1.5$ particles per de Brogile volume. The Boltzmann limit is valid for all temperatures when the number of particles per de Broglie volume is near to, or below one. The domain of validity of the newly characterized semi-dilute quantum region (\rTh{thm:mu_zero_faster}) occurs at low temperature when the number of distinguishable quantum particles per de Brogile volume is between $2$ and $3$ approximately. We see furthermore that in this study of energy content of a Fermi gas there remains at a low temperature a significant (white domain) regime in which numerical quadrature needs to be employed to obtain reasonable precision. On the other hand we see domains where two different asymptotic expansions overlap.

In \rf{fig:3-NTEP} we illustrate the regimes of validity for the energy density to pressure ratio $\varepsilon/P$, again in the plane spanned by the number of particles per de Broglie volume, and $T/m$, for $D=2,3$ dimensional systems, respectively (compare \rf{fig:2-NTT}). At high temperature the equation of state approaches $\rho=DP$ and this is properly captured by the Boltzmann approximation, which is therefore valid at high temperature even for systems with a larger number of particles in the de Broglie volume.

\begin{figure} 
\centering
 \includegraphics[width=.8\textwidth]{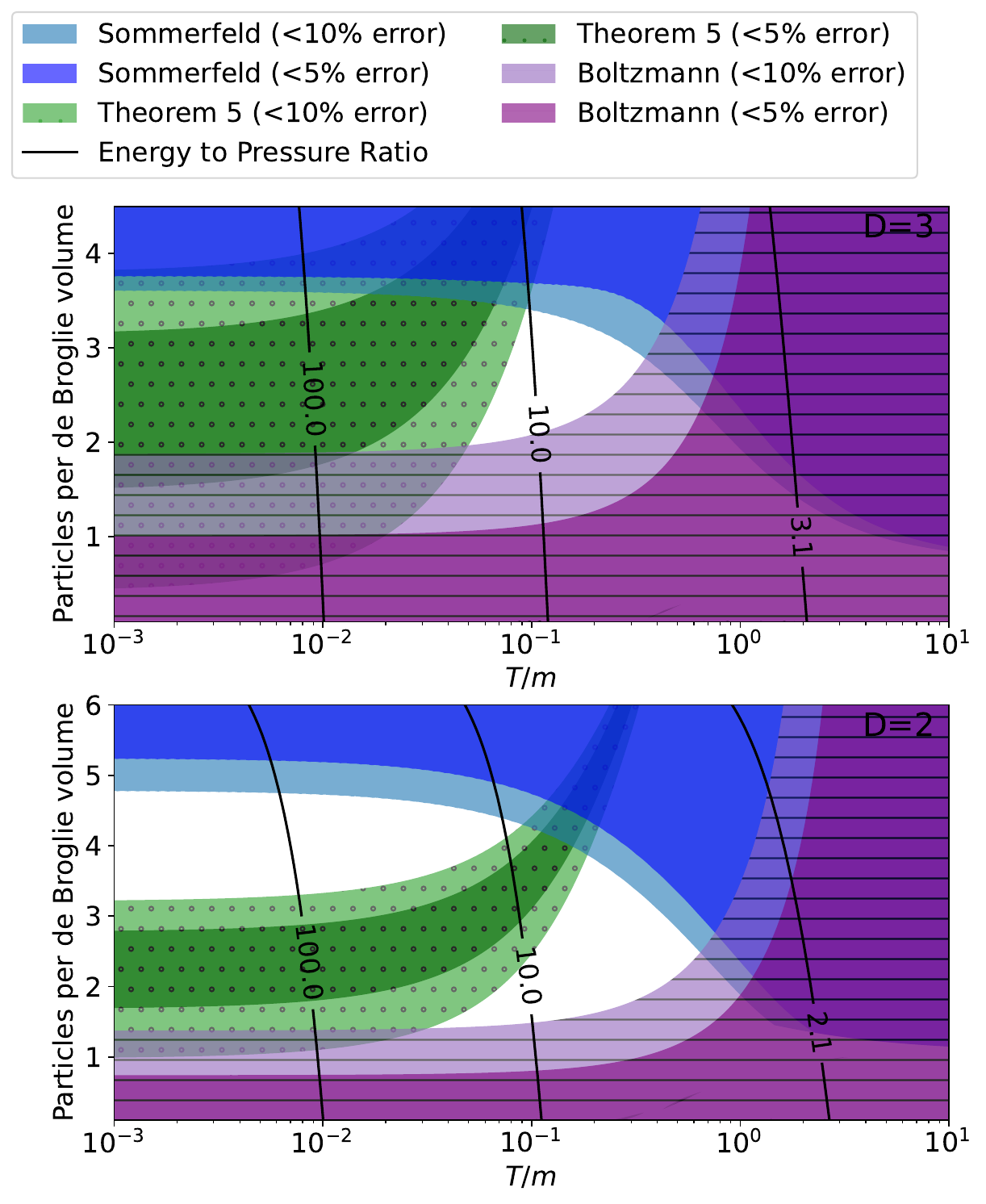}
\caption{{\xred
Study of energy density to pressure ratio $\varepsilon/P$ regimes of an ideal Fermi gas in the plane spanned by the number of particles in de Broglie volume and temperature in units of $m$, $T/m$; top frame for the three dimensional $D=3$, and bottom frame for the two dimensional $D=2$ systems: Sommerfeld expansion (blue), Boltzmann approximation (lined purple), and the new \rTh{thm:mu_zero_faster} domain (dotted green). Solid black contours show the energy to pressure ratio. The darker color applies to domains with less than 5\% error, while the lighter color allows up to 10\% error.}}\label{fig:3-NTEP}
\end{figure}

For low temperature we see in \rf{fig:3-NTEP} that the Sommerfeld regime in $D=3$ case applies when there are approximately $3.5$ or more particle per de Broglie volume while the new \rTh{thm:mu_zero_faster} domain applies in the entire gap with Boltzmann regime when the energy density to pressure ratio $\varepsilon/P$ is larger than around $10.0$, with $2\sim 3$ particles per de Broglie volume.
}

\subsection{Novel insights}\label{insights}

Our novel contributions include consideration of:
\begin{enumerate}
\item {\bf Nonrelativistic degenerate Fermi gas region (\rTh{thms:T_decay_faster}):} Application of the traditional Sommerfeld expansion in this regime requires care, as the coefficients in the expansion can become singular when $\mu\to m$. In Section \ref{sec:asymp_T_0_faster} we derive a formula that resolves those singularities so as to avoid such numerical difficulties and obtain an expansion up to a specified order. In this derivation we utilize a novel decomposition of the FD distribution into finite and zero temperature components, see Section \ref{FermiDvisit}, which we find convenient for guiding this and other derivations.
\item {\bf Degenerate relativistic Fermi gas Sommerfeld expansion (\rTh{thm:high_T_Sommerfeld_main}):} We prove that the traditional Sommerfeld expansion, which applies to the regime $T\ll m$ with $\Delta\mu$ not small, also applies to the high-temperature regime $\mu\gg T\gg m$ under appropriate assumptions.
\item {\bf Dilute quantum region (DQR) (\rTh{thm:mu_zero_faster}):} We demonstrate that the Sommerfeld expansion can fail completely when $|\mu-m|\ll T\ll m$; see the example in \rf{fig:FD_avg_expansion_comparison_Sommerfeld} below. Accurate results in this regime require the use of the new expansion that we derive in Section \ref{sec:beyondSommer}. In \rf{fig:Thm3_vs_Sommerfeld_regions} we preview the new expansion stated in \rTh{thm:mu_zero_faster}, comparing its domain of validity to that of the Sommerfeld expansion in dimensions $D=2,3$. In particular, note that the new expansion derived here covers a distinct new region in the $(T/m,\Delta\mu/m)$-plane where just a few particles are present in a de Broglie volume.
\end{enumerate}
{\xred
The energy density to pressure ratio $\varepsilon/P$ is an important physical quantity; the results shown in \rf{fig:3-NTEP} demonstrate that our new result \rTh{thm:high_T_Sommerfeld_main} provides nontrivial insights into the working of equations of state in a regime neither covered by the Sommerfeld nor by the Boltzmann approximations. }
\section{Sommerfeld-Type Asymptotic Characterization of Thermal Properties}\label{sec:Sommerfeld} 
\subsection{A new FD distribution decomposition}\label{FermiDvisit}
In this section we will explore the Sommerfeld-type asymptotic expansions of thermal averages in various physics-relevant chemical potential and temperature regimes. As a convenient tool, we will employ the following novel way to write FD distribution in the zero temperature limit $T\to0$. 

At $T=0$ the FD distribution reduces to a step function where a state $E_{i}$ is either filled or empty. For given chemical potential $\mu(T)$, we have
\begin{align}
\label{f_old}
f_\mathrm{FD}[E_{i},\mu(T),T]=\frac{1}{\displaystyle e^{\left(\frac{E_{i}-\mu}{T}\right)}+1}\,,\qquad 
\lim_{T\to0}f_\mathrm{FD}=\left\{
\begin{array}{c}
1,\quad\mathrm{for}\quad{E_{i}}<{ E_F}\\
0,\quad\mathrm{for}\quad{E_{i}}>{ E_F}
\end{array}
\right.\,.
\end{align}
At $T=0$ the energy of the last filled state is the Fermi energy ${E_\mathrm{F}=\mu(T\to0)}$, while at finite temperature one speaks of chemical potential instead. 

For the calculations in this section we find it useful to employ a novel reformulation which allows exact separation of the finite temperature behavior from the singular zero temperature limit, replacing the smooth FD distribution by the sum of three singular functions,
\begin{equation}
\frac{1}{e^{x} +1}=\Theta(-x)+\frac{1}{2}e^{-|x|}\left[\sgn(x)+\tanh(x/2)\right]\,,
\end{equation}
the first term (Heaviside function) being the $T=0$ limit and the other two describing the finite temperature residual. 

In this form the zero temperature part is explicitly separated from the finite temperature contributions. 
\begin{lemma}
 The Fermi-Dirac distribution can be decomposed into three components:
\begin{align}\label{Eq_form}
&f_\mathrm{FD}(x)=\Theta(-x)+f_\mathrm{T\neq0}(x)+\widetilde f_\mathrm{T\neq0}(x)
\end{align}
where the temperature functions are defined as
\begin{align}\label{eq:f_T_nonzero_def}
&f_\mathrm{T\neq0}=\frac{1}{2}e^{ -|x| }\mathrm{sgn}\left(x\right),\qquad
\widetilde f_\mathrm{T\neq0}=\frac{1}{2}e^{ - |x| }\tanh\left(\frac{x}{2}\right),\qquad x=\frac{E-\mu}{T}
\end{align}
and
\begin{align}
\label{NFF2}
\Theta(x)=\left\{
\begin{array}{r}
1,\quad\mathrm{for}\quad{x}>0\\
0,\quad\mathrm{for}\quad{x}<0
\end{array}\right.\,,\qquad
\sgn(x)=\left\{
\begin{array}{r}
+1,\quad\mathrm{for}\quad{x}>0\\
-1,\quad\mathrm{for}\quad{x}<0\\
\end{array}\right.\,,
\end{align}
where $\Theta(x)$ is the Heaviside step function and $\sgn(x)$ is the sign function and the equality is in the sense of distributions. In particular the values at $x=0$ are convention-dependent and do not affect our results. 
\end{lemma}

\begin{remark}\label{remark:TFdist}
To show the equivalency between the two distinct forms of the FD, we look at the three relevant regions of $x>0$, the origin $x=0$, and $x<0$. For $x>0$, \req{Eq_form} evaluates as
\begin{equation}\label{eq:xpos}
 f_\mathrm{FD}(x>0) = 0 + \frac{1}{2}e^{-x}[1+\tanh(x/2)] = (e^x + 1)^{-1}\,.
 \end{equation}
 This can be seen by using the hyperbolic formula $\tanh(x/2)=(e^x-1)/(e^x+1)$. In a similar manner, the $x<0$ region evaluates as
 \begin{equation}\label{eq:xneg}
 f_\mathrm{FD}(x<0) = 1 + \frac{1}{2}e^{x}[-1 + \tanh(x/2)] = (e^x + 1)^{-1}\,.
 \end{equation}
 As we are considering the expressions in the sense of distributions, this completes the proof. The values at $x=0$ are convention-dependent and irrelevant for our purposes, as the point $x=0$ is a set of measure zero with respect to integration.
\end{remark} 

In \rf{Fermi_Component} the solid line shows the FD distribution as a function of energy for chemical potential $\mu=0.461\MeV$, and electron mass $m=0.511\MeV$, in the top frame for temperature $T=0.02\MeV$, and in bottom frame for $T=0.2$\,MeV. Purple dashed line is the corresponding $T=0$ limit. As the temperature increases, the distribution becomes broader as a wide range of states become thermally populated. As $T$ decreases the $f_\mathrm{T\neq0}$ (blue dotted) and $\widetilde f_\mathrm{T\neq0}$ (red dash-dotted) components, see \req{eq:f_T_nonzero_def}, which characterize finite $T$ contributions become progressively more singular.

\begin{figure}
\centering
\includegraphics[width=0.72\textwidth]{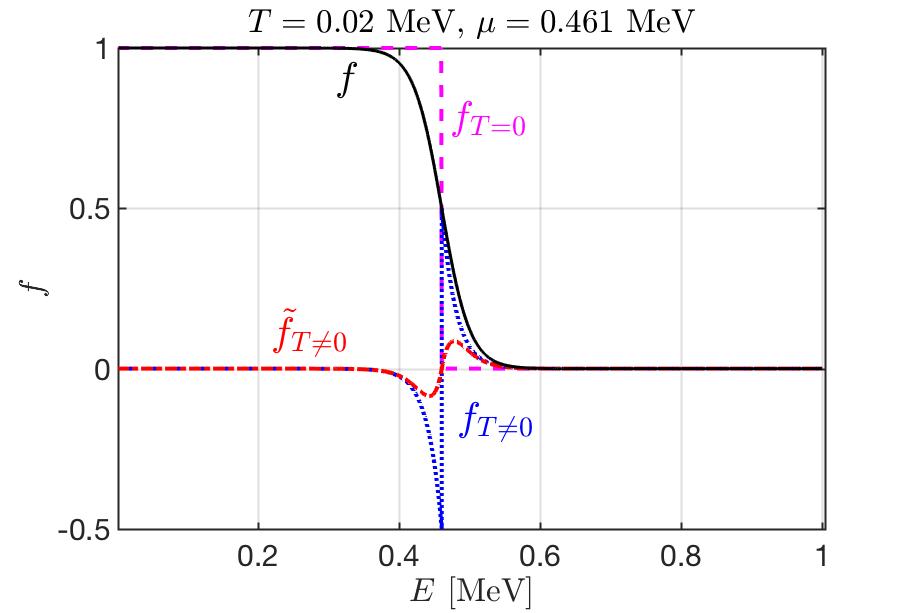}\\
\includegraphics[width=0.72\textwidth]{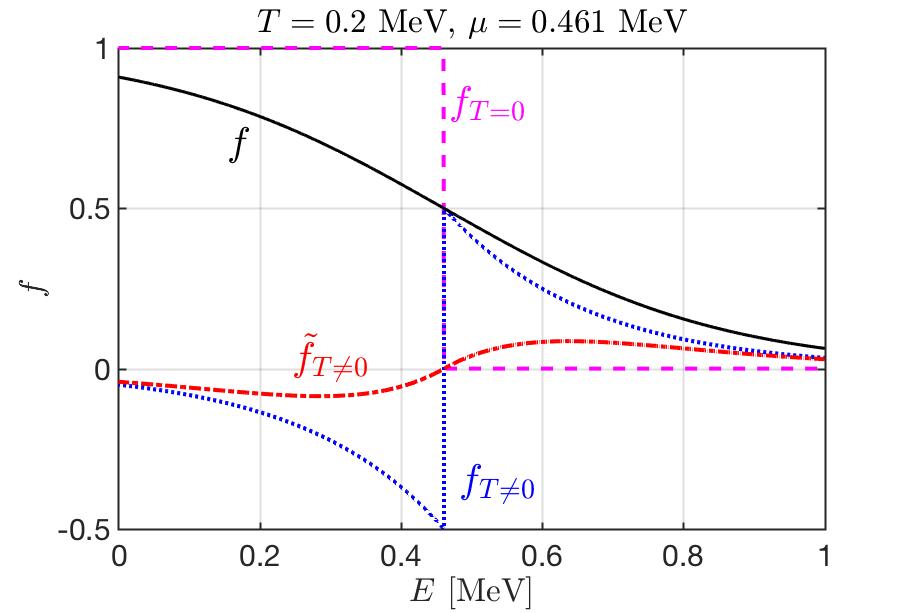}
\caption{
FD distribution (black solid line) as a function of single particle energy for chemical potential $\mu=0.461\MeV$ on top at a temperature $T=0.02\MeV$ and on bottom at $T=0.2\MeV$. The purple dashed line represents the zero temperature component $f_{\mathrm{T}=0}$; blue (dotted) and red (dash-dotted) lines represent the finite temperature components $f_\mathrm{T\neq0}$ and $\widetilde f_\mathrm{T\neq0}$ respectively, see \req{eq:f_T_nonzero_def}.
}
\label{Fermi_Component}
\end{figure}


To illustrate how this novel form of distribution can be used to address the integrals common in statistical physics especially when temperature $T\to0$, we examine the thermal average of any given function $G(p)$ with smoothness and polynomial growth bounds as $p\to \infty$. The thermal average physical quantity, denoted as $\langle G\rangle_T$, is determined by integrating over the D-dimensional phase space 
\begin{align}\label{eq:avg_G_T}
\langle G\rangle_T&\equiv\int^{\infty}_{0}\!\!\frac{d^Dp}{(2\pi)^D}\,G(p)\,f_{FD}(p)=\frac{1}{(2\pi)^D}\frac{2\pi^{D/2}}{\Gamma(D/2)}\int^{\infty}_{0}\!\!dp\,p^{D-1}\,G(p)\,f_{FD}(p).
\end{align}
By substituting our novel form of FD distribution, the physical quantity $\langle G\rangle_T$ can be expressed as the sum of two distinct components:
\begin{align}
\langle G\rangle_T=\langle G\rangle_{\Theta}+\langle G\rangle_{\Delta T}
\end{align}
 where the $\langle G\rangle_{\Theta}$ corresponds to the zero temperature limit and $\langle G\rangle_{\Delta T}$ represents the finite temperature contribution. We have
\begin{align}
&\langle G\rangle_{\Theta}=\frac{1}{(2\pi)^D}\frac{2\pi^{D/2}}{\Gamma(D/2)}\int^{\infty}_{0}\!\!dp\,p^{D-1}G(p)\Theta\left(\frac{-E+\mu}{T}\right),\qquad E=\sqrt{p^2+m^2},\\
&\langle G\rangle_{\Delta T}=\frac{1}{(2\pi)^D}\frac{2\pi^{D/2}}{\Gamma(D/2)}\int^{\infty}_{0}\!\!dp\,p^{D-1}\,G(p)\,\left[f_\mathrm{T\neq0}(p)+\widetilde f_\mathrm{T\neq0}(p)\right].\label{G_deltaT}
\end{align}

By employing our FD decomposition, in the following sections we obtain asymptotic expansions of the finite temperature contribution $\langle G\rangle_{\Delta T}$ in different physical regimes of interest. Our novel decomposition guides the asymptotic analysis of $\langle G\rangle_{\Delta T}$ by naturally reorganizing the integrand into a difference of integrals with exponentially decaying integrands. This allows us to utilize asymptotic analysis techniques to extract leading and subleading order behavior as $T\to 0$ and to analytically account for the exact cancellation of certain terms, which otherwise cause problems in numerical computations. 





\subsection{Traditional Sommerfeld Regime}\label{sec:traditional_Sommerfeld}
We now show that the decomposition \req{Eq_form} naturally motivates certain steps in the derivation of the traditional Sommerfeld expansion, see, e.g., Eq. (58.1) on page 170 of \cite{landau2013statistical}, an asymptotic expansion for the difference between a $D$-dimensional thermal average at finite and zero temperature as $T\to 0$ in the regime where $\mu>m$. In this derivation we will be particularly attentive to the location of singularities as well as precisely bounding error terms in the asymptotic expansion. While these details do not play a critical role in obtaining the correct form of the traditional Sommerfeld expansion, they will be important in our new results later in this work and so we wish to review these methods in a more familiar setting first.

Here we will assume that $\mu>m$, $k>1$ is odd, and the density of states $H(E)$ is polynomially bounded on $[m,\infty)$ and is $C^k$ on a neighborhood of $\mu$. We will obtain an asymptotic expansion of 
\begin{align}\label{eq:traditional_sommerfeld_I}
 I\equiv\int_m^\infty dE\, H(E) f_{FD}(E)
\end{align}
as $T\to 0$. First we note that, starting from $G(p)$ in \req{G_deltaT}, the corresponding $H(E)$ is given by
\begin{align}
 H(E)= \frac{1}{(2\pi)^D}\frac{2\pi^{D/2}}{\Gamma(D/2)} E(E^2-m^2)^{(D-2)/2} G(\sqrt{E^2-m^2}) 
\end{align}
and this $H$ has the required properties if $D\geq 2$ and $G$ is polynomially bounded on $[0,\infty)$ and is $C^k$ on a neighborhood of $\sqrt{\mu^2-m^2}$ (recall that we are assuming $\mu>m$, which allows us to avoid the singularity at $E=m$).

We start by using the decomposition \req{Eq_form} to break the integral into zero and nonzero temperature contributions
\begin{align}\label{eq:I_split}
 I= \int_m^\mu dE\, H(E) +\int_m^\infty dE\, H(E) \left[f_{T\neq 0}(E)+\widetilde{f}_{T\neq 0}(E)\right]\,.
\end{align}
Focusing on the non-zero temperature contribution in \req{eq:I_split}, we break the second integral into two terms as motivated by the discontinuity of \req{eq:f_T_nonzero_def} at $E=\mu$ and then simplify the integrands using \req{eq:xpos} and \req{eq:xneg} as follows:
\begin{align}
 &\int_m^\infty dE\,H(E) \left[f_{T\neq 0}(E)+\widetilde{f}_{T\neq 0}(E)\right]\\
=& \int_\mu^\infty dE\,H(E) \left[f_{T\neq 0}(E)+\widetilde{f}_{T\neq 0}(E)\right]+ \int_m^\mu dE\,H(E) \left[f_{T\neq 0}(E)+\widetilde{f}_{T\neq 0}(E)\right]\notag\\ 
=&\int_\mu^\infty dE\, H(E) \frac{1}{1+e^{(E-\mu)/T}}-\int_m^\mu dE\,H(E) \frac{1}{1+e^{-(E-\mu)/T}}\,.\label{eq:traditional_sommerfeld_2_integral}
\end{align}
The exponentially decaying factors in the above integrands imply that reducing the domains of integration to the intervals $[\mu,\mu+\delta]$ and $[\mu-\delta,\mu]$ respectively for any choice of $\delta>0$ incurs an $\mathcal{O}(e^{-c/T})$ error term for some $c>0$; the assumptions on $H$ imply we can choose $\delta$ so that $H$ is $C^k$ on $[\mu-\delta,\mu+\delta]$. To see that the error does in fact have the claimed behavior, use the polynomial boundedness $|H(E)|\leq CE^q$ for some $C,q>0$ and change variables to $z=(E-\mu)/T$ to compute
\begin{equation}\label{eq:change_limits_int_exponential_decay}
 \begin{split}
 &\left|\int_{\mu+\delta}^\infty dE\, H(E)\frac{1}{1+e^{(E-\mu)/T}}\right|\\
 &\leq C\int_{\mu+\delta}^\infty dE\,E^q e^{-(E-\mu)/T} = CT\int_{\delta/T}^\infty dz\,(Tz+\mu)^q e^{-z}\\
 &\leq 2^qCT\left(T^q \int_{\delta/T}^\infty dz\,z^q e^{-z}+\mu^q \int_{\delta/T}^\infty dz\,e^{-z}\right) =\mathcal{O}\left(e^{-c/T}\right)
 \end{split}
\end{equation}
for some $c>0$. The second integral in \req{eq:traditional_sommerfeld_2_integral} is handled similarly. The $\mathcal{O}(e^{-c/T})$ error is dominated by $\mathcal{O}(T^{r})$ for any $r$ and so the exact value of $c$ will be seen to be irrelevant for our purposes.

The previous analysis then implies 
\begin{align}
 \req{eq:traditional_sommerfeld_2_integral} =&\int_\mu^{\mu+\delta} \!dE\, H(E) \frac{1}{1+e^{(E-\mu)/T}}-\int_{\mu-\delta}^\mu \!dE\,H(E) \frac{1}{1+e^{-(E-\mu)/T}}+\mathcal{O}\left(e^{-c/T}\right)\,,\notag\\
 =&T\int_0^{\delta/T} dz\, \left[H(\mu+Tz) -H(\mu-Tz)\right] \frac{1}{1+e^{z}}+\mathcal{O}\left(e^{-c/T}\right)\,,\label{eq:traditional_sommerfeld_2_integral_combined} 
\end{align}
where to obtain the last line we changed variables to $z=(E-\mu)/T$ in the first integral and $z=-(E-\mu)/T$ in the second. The domain of integration is now such that $H$ is $C^k$ at $\mu\pm Tz$ for all $z\in[0,\delta/T]$ and therefore we can compute a Taylor expansion up to order $k$ with remainder,
\begin{align}
 &H(\mu+Tz)- H(\mu-Tz)=\sum_{n=1,n\text{ odd}}^{k-2} 2H^{(n)}(\mu)\frac{(Tz)^n}{n!}+R_k(z)\,,\\
 &R_k(z)\equiv (Tz)^k \int_0^1 ds\,\frac{(1-s)^{k-1}}{(k-1)!} \left[H^{(k)}(\mu+sTz)+H^{(k)}(\mu-sTz)\right]\,,
\end{align}
and $|H^{(k)}(\mu\pm sTz)|\leq C_k\equiv \sup_{y\in[\mu-\delta,\mu+\delta]}|H^{(k)}(y)|<\infty$ for all $z\in[0,\delta/T]$, $s\in[0,1]$ (it is finite due to continuity of $H^{(k)}$ on the the compact interval $[\mu-\delta,\mu+\delta]$). Therefore the integral of the remainder term can be bounded as follows
\begin{align}
 \left|T\int_0^{\delta/T} R_k(z)\frac{1}{1+e^z}\right|
 \leq&\frac{2C_k}{k!} T^{k+1}\int_0^{\delta/T}dz\, z^k e^{-z}=\mathcal{O}\left(T^{k+1}\right)\,.
\end{align}
Combining this with \req{eq:traditional_sommerfeld_2_integral_combined} we obtain
\begin{align}
 \req{eq:traditional_sommerfeld_2_integral} 
 =&\sum_{n=1,n\text{ odd}}^{k-2} 2\frac{T^{n+1}}{n!} H^{(n)}(\mu)\int_0^{\delta/T} dz\,\frac{z^n}{1+e^{z}}+\mathcal{O}\left(T^{k+1}\right)+\mathcal{O}\left(e^{-c/T}\right)\,.
\end{align}
Similarly to the case \req{eq:change_limits_int_exponential_decay}, the exponential decay factor in the integrand implies that we can change the domains of integration to $[0,\infty)$ at the cost of an $\mathcal{O}(e^{-c/T})$ error term. Using this and then the integral formula
\begin{align}\label{eq:FD_power_integrals}
 \int_0^\infty \frac{x^{\nu-1}}{e^{ x}+1}dx=\begin{cases}
 \ln(2)\,, &\quad\mathrm{for}\quad \nu=1\\
 (1-2^{1-\nu})\Gamma(\nu)\zeta(\nu)\,, &\quad\mathrm{for}\quad \nu > 1
 \end{cases}
\end{align}
(see 3.411.3 on page 353 of \cite{Gradshteyn:1943cpj}) we obtain 
\begin{align}
 \req{eq:traditional_sommerfeld_2_integral} 
 =&\sum_{n=1,n\text{ odd}}^{k-2} 2(1-2^{-n})\zeta(n+1)T^{n+1} H^{(n)}(\mu) +\mathcal{O}\left(T^{k+1}\right)\,.
\end{align}
Substituting this into \req{eq:traditional_sommerfeld_I} we obtain the traditional Sommerfeld expansion
\begin{align}
 &\int_m^\infty dE\, H(E) f_{FD}(E)\notag\\
 = &\int_m^\mu dE\, H(E)+\sum_{n=1,n\text{ odd}}^{k-2} 2(1-2^{-n})\zeta(n+1)T^{n+1} H^{(n)}(\mu) +\mathcal{O}(T^{k+1})\,.\label{eq:traditional_Sommerfeld_final}
 \end{align}
 We emphasize that the expansion \req{eq:traditional_Sommerfeld_final} considers $\mu$ to be fixed and so the implied constant in the $\mathcal{O}(T^{k+1})$ error term has nontrivial $\mu$-dependence. This fact causes the failure of the traditional Sommerfeld expansion in certain cases where one is considering simultaneous $\mu$ and $T$ limits. This is not simply a mathematical consideration, as simultaneous limits are required to maintain appropriate physical parameters as $T\to 0$, e.g., the moderate density regime shown in Figure \ref{fig:Thm3_vs_Sommerfeld_regions}. In the remainder of this work we derive several novel asymptotic expansions of FD integrals by precisely considering different $\mu$, $T$ limits. The expansions in Sections \ref{sec:asymp_T_0_faster} and \ref{Section:HighTempSommerfeld} can still be considered to be of Sommerfeld type, while the regime considered in Section \ref{sec:beyondSommer} leads to a wholly novel expansion.

\subsection{Sommerfeld Expansion in the Nonrelativistic Degenerate Fermi Gas Regime $T\ll \mu-m\ll m$}\label{sec:asymp_T_0_faster}
In this section we use the decomposition \req{Eq_form} to further investigate the Sommerfeld expansion in the limit where $T\ll\mu-m\ll m$. While in essence the traditional Sommerfeld expansion applies to this regime, a naive use of \req{eq:traditional_Sommerfeld_final} often results in one encountering numerical difficulties due to singular behavior of various coefficients in the limit $\mu\to m$. In this section we derive a formula that precisely accounts for these singularities; this formula makes it straightforward to compute the coefficients in the expansion up to any desired order via a computer algebra system.

More specifically, in this section we will assume $D\geq 2$ and
\begin{align}\label{eq:T_mu_relation}
 \mu=m+\Delta\mu\,, \,\Delta\mu>0 \,\, \text{ and } \,\, \widetilde{T}=d\Delta\widetilde{\mu}^{1+\gamma}\,, \,\,d,\gamma>0\,,
\end{align}
where $\widetilde{T}\equiv T/m$ and $\Delta\widetilde{\mu}\equiv \Delta\mu/m$; these scalings imply that $T\ll\mu-m\ll m$ as $\Delta\widetilde{\mu}\to 0$, matching the stated goal here. We let $G(p)$, $p\in[0,\infty)$, be a $C^k$ function whose zeroth through $k$'th derivatives are polynomially bounded and will study $\langle G\rangle_{\Delta T}$ in the limit $\Delta\widetilde{\mu}\to 0^+$. The final result is found in \rTh{thms:T_decay_faster} below. To proceed with the derivation, first change variables in \req{G_deltaT} to $E=\sqrt{p^2+m^2}$ and then use \req{eq:xpos} and \req{eq:xneg}:
\begin{align}
 &\int_0^\infty dp\, p^{D-1} G(p)(f_{T\neq 0}+\widetilde{f}_{T\neq 0})\notag\\
 =&\int_m^\infty dE \,E(E^2-m^2)^{(D-2)/2} G\left(\sqrt{E^2-m^2}\right)(f_{T\neq 0}+\widetilde{f}_{T\neq 0})\notag\\
 =&\int_{\mu}^\infty dE\, E(E^2-m^2)^{(D-2)/2} G\left(\sqrt{E^2-m^2}\right)
 \frac{1}{1+e^{(E-\mu)/T}}\label{eq:2_int_decomp_T_faster} \\
 &-\int_m^{\mu} dE \,E(E^2-m^2)^{(D-2)/2} G\left(\sqrt{E^2-m^2}\right)\frac{1}{1+e^{(\mu-E)/T}} \,.\notag
\end{align}
Next consider the first integral in \req{eq:2_int_decomp_T_faster}. In general, its integrand will have a singularity at $E=m$. As the lower limit of integration approaches $m$ in the regime under consideration, it is necessary to carefully resolve that singularity. We have chosen to begin with the integral expressed in terms of $p$ because it simplifies this analysis, due to $E=\sqrt{p^2+m^2}$ being is a smooth function of $p$ but not vice versa. Thus if one begins with a smooth density of states $H(E)$, the corresponding $G(p)$ will also be smooth but the reverse is not true. We note that if one is interested in $G(p)$ that is not smooth at $p=0$ then the following derivation can be easily modified to account for a power-law singularity of $G$ at $0$.

To derive an expansion that correctly handles the singular behavior of the integrands in \req{eq:2_int_decomp_T_faster} at $E=m$, we first change variables to $z=(E-m)/m$ to obtain
\begin{align}
&\int_{\mu}^\infty dE\, E(E^2-m^2)^{(D-2)/2} G\left(\sqrt{E^2-m^2}\right)
 \frac{1}{1+e^{(E-\mu)/T}}\notag \\
 =&m^{D}\int_{\Delta\widetilde{\mu}}^\infty dz\, (z+1)[(z+1)^2-1]^{(D-2)/2} G\left(m\sqrt{(z+1)^2-1}\right)
 \frac{1}{1+e^{(z-\Delta\widetilde{\mu})/\widetilde{T}}} \notag\\
 =&m^{D}\int_{\Delta\widetilde{\mu}}^\infty dz\, (z+1)( z+2)^{(D-2)/2} G\left(m\sqrt{z}\sqrt{z+2}\right)
 \frac{z^{(D-2)/2}}{1+e^{(z-\Delta\widetilde{\mu})/\widetilde{T}}} \,.\label{eq:T_faster_int_1_change_vars}
\end{align}
We have separated out the (possible) singularity coming from $z^{(D-2)/2}$ but there still remains a potential singularity at $z=0$ coming from the argument of $G$. The term involving $G$ does not always have a Taylor expansion, but it does have a more general asymptotic expansion which we can obtain as follows. Define 
\begin{align}\label{eq:F_y_m_def}
 F(y,m)\equiv (y^2+1)( y^2+2)^{(D-2)/2} G\left(my\sqrt{y^2+2}\right)\,.
\end{align}
We have assumed that $G$ is $C^k$ on $[0,\infty)$ with polynomially bounded zeroth through $k$'th derivatives and therefore $y\mapsto F(y,m)$ also has these properties. Hence we can use Taylor's theorem with remainder to write
\begin{align}\label{eq:F_y_m_Taylor}
F(y,m)=&\sum_{n=0}^{k-1} a_n(m)y^n +R_k(y,m)\,, \,\,\,a_n(m)=\frac{1}{n!}\partial_y^n F(0,m)\,,
\end{align}
where the remainder term is given by
\begin{align}
R_k(y,m)=y^k \int_0^1\frac{(1-s)^{k-1}}{(k-1)!}\partial_y^k F(sy,m)ds
\end{align}
and can be bounded via
\begin{align}\label{eq:R_k_poly_bound}
|R_k(y,m)|\leq y^k[\alpha_k(m)+\beta_k(m)y^{q_k}]
\end{align}
for some coefficients $\alpha_k,\beta_k\geq 0$ and power $q_k\geq 0$ due to the polynomial boundedness of $\partial_y^k F$.

With this we have
\begin{align}
\req{eq:T_faster_int_1_change_vars} =&m^{D}\!\int_{\Delta\widetilde{\mu}}^\infty \!dz\, F(\sqrt{z},m) \frac{z^{(D-2)/2}}{1+e^{(z-\Delta\widetilde{\mu})/\widetilde{T}}} \\
=& \sum_{n=0}^{k-1} a_n(m)m^{D}\!\!\int_{\Delta\widetilde{\mu}}^\infty \!dz\, \frac{z^{(n+D-2)/2}}{1+e^{(z-\Delta\widetilde{\mu})/\widetilde{T}}}+m^{D}\!\!\int_{\Delta\widetilde{\mu}}^\infty\! dz\,R_k(\sqrt{z},m)\frac{z^{(D-2)/2}}{1+e^{(z-\Delta\widetilde{\mu})/\widetilde{T}}}\,.\notag
\end{align}
The integral of the remainder term can be bounded by using \req{eq:R_k_poly_bound} as follows
\begin{align}
 &\left|m^{D}\int_{\Delta\widetilde{\mu}}^\infty dz\,R_k(\sqrt{z},m)\frac{z^{(D-2)/2}}{1+e^{(z-\Delta\widetilde{\mu})/\widetilde{T}}}\right|\\
 \leq&m^{D}\int_{\Delta\widetilde{\mu}}^\infty dz\,z^{k/2}[\alpha_k(m)+\beta_k(m)z^{q_k/2}]z^{(D-2)/2}e^{-(z-\Delta\widetilde{\mu})/\widetilde{T}}\notag\\
=&d\alpha_k(m)m^{D}\Delta\widetilde{\mu}^{(k+D)/2+\gamma}\int_{0}^\infty dx\,(1+d\Delta\widetilde{\mu}^{\gamma}x)^{(k+D-2)/2}e^{-x}\notag\\
&+d\beta_k(m)m^{D}\Delta\widetilde{\mu}^{(k+q_k+D)/2+\gamma}\int_{0}^\infty dx\,(1+d\Delta\widetilde{\mu}^{\gamma}x)^{(k+q_k+D-2)/2}e^{-x}\notag\\ 
=&\mathcal{O}\left(\Delta\widetilde{\mu}^{(k+D)/2+\gamma}\right)\quad \text{as}\quad\Delta\widetilde{\mu}\to 0^+\,,\label{eq:T_faster_remainder_bound}
\end{align}
where we changed variables to $x=(z-\Delta\widetilde{\mu})/\widetilde{T}$ and used \req{eq:T_mu_relation}. Making this same change of variables in the remaining terms, we therefore obtain
\begin{align}\label{eq:T_faster_int_1_change_vars_expanded1}
\req{eq:T_faster_int_1_change_vars} =& \sum_{n=0}^{k-1} da_n(m)m^{D}\Delta\widetilde{\mu}^{(n+D)/2+\gamma}\!\int_{0}^\infty\! dx\, \frac{(1+d\Delta\widetilde{\mu}^{\gamma}x)^{(n+D-2)/2}}{1+e^{x}}\\
&+\mathcal{O}\left(\Delta\widetilde{\mu}^{(k+D)/2+\gamma}\right)\,.\notag
\end{align}
These remaining integrals are smooth as functions of $d\Delta\widetilde{\mu}^\gamma$ and can be Taylor expanded to order $k-1$ by differentiating under the integral, with remainder $\mathcal{R}_k(C,x)$:
\begin{align}\label{eq:aux_int_expansion1}
&\int_0^\infty dx\, \frac{(1+C x)^\nu}{1+e^x} = \int_0^\infty dx\left(\sum_{j=0}^{k-1} \frac{1}{j!} \left.\frac{\partial^j}{\partial C^j}\right|_{C=0}\!\!\! \frac{(1+Cx)^\nu}{1 + e^x} C^j + \mathcal{R}_k(C,x)\right)\\
=&\sum_{j=0}^{k-1}\frac{1}{j!} \left(\prod_{\ell=0}^{j-1}(\nu-\ell) \right)C^j\int_0^\infty dx\frac{x^j}{1+e^x} \notag\\
&+C^k\int_0^\infty dx \frac{x^k}{1+e^x} \int_0^1 ds\frac{(1-s)^{k-1}}{(k-1)!} \left(\prod_{\ell=0}^{k-1}(\nu-\ell)\right)(1+sCx)^{\nu-k}\notag\\
=&\ln(2)+\sum_{j=1}^{k-1}\frac{1}{j!} \left(\prod_{\ell=0}^{j-1}(\nu-\ell) \right)C^j (1 - 2^{-j}) \Gamma(j+1) \zeta(j+1) +\mathcal{O}(C^k)\,, \notag
\end{align}
where we used the integral formula \req{eq:FD_power_integrals}. Now we apply this to the terms in \req{eq:T_faster_int_1_change_vars_expanded1} with $C=d\Delta\widetilde{\mu}^\gamma$ and $\nu=(n+D-2)/2$; note that for each $n$ we must expand \req{eq:aux_int_expansion1} up to order $k_n\equiv\lceil(k-n)/(2\gamma)\rceil$ so that the the remainder term in \req{eq:aux_int_expansion1} is of the same or higher order than the remainder in \req{eq:T_faster_remainder_bound} for each $n$. Doing this yields
\begin{align}\label{eq:T_faster_int_1_change_vars_expanded2}
\req{eq:T_faster_int_1_change_vars} =& \sum_{n=0}^{k-1} da_n(m)m^{D}\Delta\widetilde{\mu}^{(n+D)/2+\gamma}\bigg[
\ln(2)\\
&+\sum_{j=1}^{k_n-1}d^j\Delta\widetilde{\mu}^{j\gamma} \left(\prod_{\ell=0}^{j-1}[(n+D-2)/2-\ell] \right) (1 - 2^{-j}) \zeta(j+1)\bigg]\notag
\\
&+\mathcal{O}\left(\Delta\widetilde{\mu}^{(k+D)/2+\gamma}\right)\,.\notag
\end{align}
This completes the asymptotic expansion of the first integral in \req{eq:2_int_decomp_T_faster}.

Now consider the second integral in \req{eq:2_int_decomp_T_faster}. Again we change variables to $z=(E-m)/m$ and make use of \req{eq:F_y_m_def} and \req{eq:F_y_m_Taylor}:
\begin{align}\label{eq:T_faster_int_2_change_vars} 
 &\int_m^{\mu} dE \,E(E^2-m^2)^{(D-2)/2} G\left(\sqrt{E^2-m^2}\right)\frac{1}{1+e^{(\mu-E)/T}} \\\notag
 =&m^{D}\int_0^{\Delta\widetilde{\mu}} dz \, F(\sqrt{z},m)\frac{z^{(D-2)/2}}{1+e^{(\Delta\widetilde{\mu}-z)/\widetilde{T}}}\notag\\
 =&\sum_{n=0}^{k-1} a_n(m)m^{D}\int_0^{\Delta\widetilde{\mu}} dz \,\frac{z^{(n+D-2)/2}}{1+e^{(\Delta\widetilde{\mu}-z)/\widetilde{T}}} +m^{D}\int_0^{\Delta\widetilde{\mu}} dz \,R_k(\sqrt{z},m)
\frac{z^{(D-2)/2}}{1+e^{(\Delta\widetilde{\mu}-z)/\widetilde{T}}}\notag\,.
\end{align}
We bound the integral of the remainder by again using \req{eq:R_k_poly_bound}
\begin{align}
 &\left|m^{D}\int_0^{\Delta\widetilde{\mu}} dz \,R_k(\sqrt{z},m)
\frac{z^{(D-2)/2}}{1+e^{(\Delta\widetilde{\mu}-z)/\widetilde{T}}}\right|\\
\leq&m^{D}\int_0^{\Delta\widetilde{\mu}} dz \,\left[\alpha_k(m)+\beta_k(m)z^{q_k/2}\right]z^{(k+D-2)/2}e^{-(\Delta\widetilde{\mu}-z)/\widetilde{T}}\notag\\
\leq &m^{D}\widetilde{T}\left[\alpha_k(m)\Delta\widetilde{\mu}^{(k+D-2)/2}+\beta_k(m)\Delta\widetilde{\mu}^{(k+q_k+D-2)/2}\right]\notag\\
=&\mathcal{O}\left(\Delta\widetilde{\mu}^{(k+D)/2+\gamma}\right)\,.\notag
\end{align}
Therefore 
\begin{align} \label{eq:T_faster_int_2_expanded1}
 \req{eq:T_faster_int_2_change_vars} =&\sum_{n=0}^{k-1} da_n(m)m^{D}\Delta\widetilde{\mu}^{(n+D)/2+\gamma}\int^{\Delta\widetilde{\mu}/\widetilde{T}}_{0} du \,\frac{(1-\widetilde{T}u/\Delta\widetilde{\mu})^{(n+D-2)/2}}{1+e^{u}} \\
 &+\mathcal{O}(\Delta\widetilde{\mu}^{(k+D)/2+\gamma})\notag\,,
\end{align}
where we changed variables to $u=(\Delta\widetilde{\mu}-z)/\widetilde{T}$. The remaining integrals can be expanded using the following computation for $C>0$: 
\begin{align}
&\int_0^{1/C} du \,\frac{(1-Cu)^\nu}{1+e^u} = \int_0^{1/C} du\left(\sum_{j=0}^{k-1} \frac{1}{j!} \left. \frac{\partial^j}{\partial C^j}\right|_{C=0}!\!\!\frac{(1-Cu)^\nu}{1 + e^u} C^j + \mathcal{R}_k(C,u)\right)\\
=&\sum_{j=0}^{k-1}\frac{C^j}{j!}(-1)^j\left(\prod_{\ell=0}^{j-1}(\nu-\ell) \right)\int_0^{1/C} du \,\frac{u^j}{1+e^u} \notag\\
&+ C^k (-1)^k\left(\prod_{\ell=0}^{k-1}(\nu-\ell)\right)\int_0^{1/C} du \frac{1}{1+e^u} u^k\int_0^1ds\,\frac{(1-s)^{k-1}}{(k-1)!} (1-Csu)^{\nu-k}\,.\notag
\end{align}
When $\nu\geq k$ we have $(1-Csu)^{\nu-k}\leq 1$ and so it is straightforward to see that the integral of the remainder $\mathcal{R}_k(C,u)$ is $\mathcal{O}(C^k)$ as $C\to 0$. When $\nu<k$ we have $(1-Csu)^{\nu-k}\leq (1-s)^{\nu-k}$ and so the remainder can be bounded as follows
\begin{align}
&\left|C^k (-1)^k\left(\prod_{\ell=0}^{k-1}(\nu-\ell)\right)\int_0^{1/C} du \frac{1}{1+e^u} u^k\int_0^1ds\,\frac{(1-s)^{k-1}}{(k-1)!} (1-Csu)^{\nu-k}\right|\notag\\
\leq&C^k\left| \left(\prod_{\ell=0}^{k-1}(\nu-\ell)\right)\right|\int_0^{1/C} du \frac{1}{1+e^u} u^k\int_0^1ds\,\frac{(1-s)^{\nu-1}}{(k-1)!} =\mathcal{O}(C^k)
\end{align}
for all $\nu>0$. Therefore we have shown
\begin{align}
\int_0^{1/C} du \,\frac{(1-Cu)^\nu}{1+e^u}
=\sum_{j=0}^{k-1}\frac{C^j}{j!}(-1)^j\left(\prod_{\ell=0}^{j-1}(\nu-\ell) \right)\int_0^{1/C} du \,\frac{u^j}{1+e^u} +\mathcal{O}(C^k)
\end{align}
(note that this trivially holds for $\nu=0$ as well). Also note that the exponential decay of the integrand implies 
\begin{align}
\int_{1/C}^\infty du\, \frac{u^j}{1+e^u}=\mathcal{O}(C^k)
\end{align} for any $k$ and hence we are free to change the upper limits of integration to $\infty$ without changing the order of the error, thereby yielding:
\begin{align}
\int_0^{1/C} du \,\frac{(1-Cu)^\nu}{1+e^u}
=\sum_{j=0}^{k-1}\frac{C^j}{j!}(-1)^j\left(\prod_{\ell=0}^{j-1}(\nu-\ell) \right)\int_0^{\infty} du \,\frac{u^j}{1+e^u} +\mathcal{O}(C^k)\,.
\end{align}
For each $n$ we use this expansion up to order $k_n= \lceil(k-n)/(2\gamma)\rceil$ with $C\equiv \widetilde{T}/\Delta\widetilde{\mu}=d\Delta\widetilde{\mu}^{\gamma}$ and $\nu=(n+D-2)/2$ with $n\geq 0$, $D\geq 2$, along with the integral formulas 
\req{eq:FD_power_integrals}. Substituting these into \req{eq:T_faster_int_2_expanded1} gives
\begin{align} \label{eq:T_faster_int_2_expanded2}
 \req{eq:T_faster_int_2_change_vars} =&\sum_{n=0}^{k-1} da_n(m)m^{D}\Delta\widetilde{\mu}^{(n+D)/2+\gamma}\bigg[\ln(2)\\
 &+\sum_{j=1}^{k_n-1}(-1)^jd^j\Delta\widetilde{\mu}^{j\gamma}\left(\prod_{\ell=0}^{j-1}[(n+D-2)/2-\ell] \right)(1-2^{-j})\zeta(j+1) \bigg] \notag\\
 &+\mathcal{O}\left(\Delta\widetilde{\mu}^{(k+D)/2+\gamma}\right)\notag\,,
\end{align}
This completes the asymptotic expansion of the second integral in \req{eq:2_int_decomp_T_faster}.

Finally, subtracting \req{eq:T_faster_int_2_expanded2} from \req{eq:T_faster_int_1_change_vars_expanded2} and canceling the shared terms (in particular, they cancel at leading order) we obtain the following asymptotic expansion.

\begin{theorem}\label{thms:T_decay_faster}
Let $D\geq 2$, $\mu=m+\Delta\mu$, $\Delta\mu>0$ and $\widetilde{T}=d\Delta\widetilde{\mu}^{1+\gamma}$ {\xred with} $d,\gamma>0$. Let $k\in\mathbb{Z}^+$ and suppose $G(p)$ is a $C^k$ function on $p\in[0,\infty)$ whose zeroth through $k$'th derivatives are polynomially bounded. Then
\begin{align}
 &\int_0^\infty dp\, p^{D-1} G(p)(f_{T\neq 0}+\widetilde{f}_{T\neq 0})\label{eq:T_faster_orig_integral}\\
 =& \sum_{n=0}^{k-1} da_n(m)m^{D}\Delta\widetilde{\mu}^{(n+D)/2+\gamma}\label{eq:T_faster_expansion_final}\\
 &\times\sum_{j=1,\mathrm{odd}}^{\lceil(k-n)/(2\gamma)\rceil-1}
 2d^j\Delta\widetilde{\mu}^{j\gamma}\left(\prod_{\ell=0}^{j-1}[(n+D-2)/2-\ell] \right)(1-2^{-j})\zeta(j+1) \notag\\
 &+\mathcal{O}\left(\Delta\widetilde{\mu}^{(k+D)/2+\gamma}\right)\notag
\end{align} 
as $\Delta\widetilde{\mu}\to 0^+$, where $\Delta\widetilde{\mu}=(\mu-m)/m$, $\widetilde{T}=T/m$, and
\begin{align}\label{eq:an_def}
a_n(m)=\frac{1}{n!}\partial_y^n|_{y=0}\left[ (y^2+1)( y^2+2)^{(D-2)/2} G\left(my\sqrt{y^2+2}\right)\right]\,.
\end{align}
\end{theorem}
We provide explicit formulas for the first few $a_n$'s below:
\begin{align}\label{eq:first_few_a_ns}
 &a_0(m)=2^{(D-2)/2} G(0)\,,\\
 &a_1(m)=2^{(D-1)/2}mG^\prime(0)\,,\notag\\
 &a_2(m)=2^{(D-6)/2}\left[(D+2)G(0)+4m^2 G^{\prime\prime}(0)\right]\,,\notag\\
 &a_3(m)=2^{(D - 5)/2} \left((D + 3) m G^\prime(0) + \frac{4}{3}m^3 G^{\prime\prime\prime}(0) \right)\,,\notag
\end{align}
where the primes denote derivatives of $G(p)$ with respect to $p$. In dimension $D=3$ the leading order term in \req{eq:T_faster_expansion_final} will be determined by the first $n$ for which $a_n(m)\neq 0$, which results in a term that scales with $\Delta\widetilde{\mu}^{(n+3)/2+2\gamma}$. In \rf{fig:T_faster_expansion_comparison} we show a comparison between the expansion \req{eq:T_faster_expansion_final} and numerical integration of \req{eq:T_faster_orig_integral}, demonstrating the effectiveness of the expansion as well as error scaling that matches the result \req{eq:T_faster_expansion_final}.

\begin{figure} [h]
\centering
\includegraphics[width=0.49\textwidth]{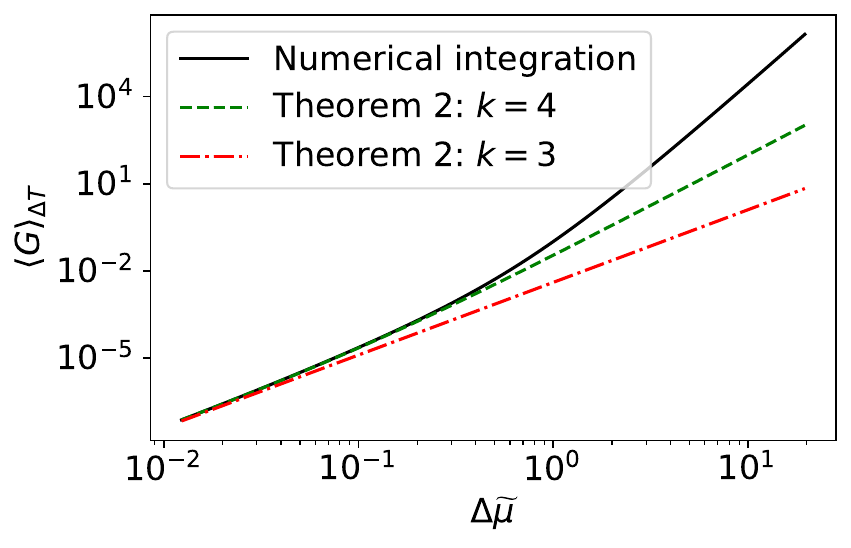}
\includegraphics[width=0.49\textwidth]{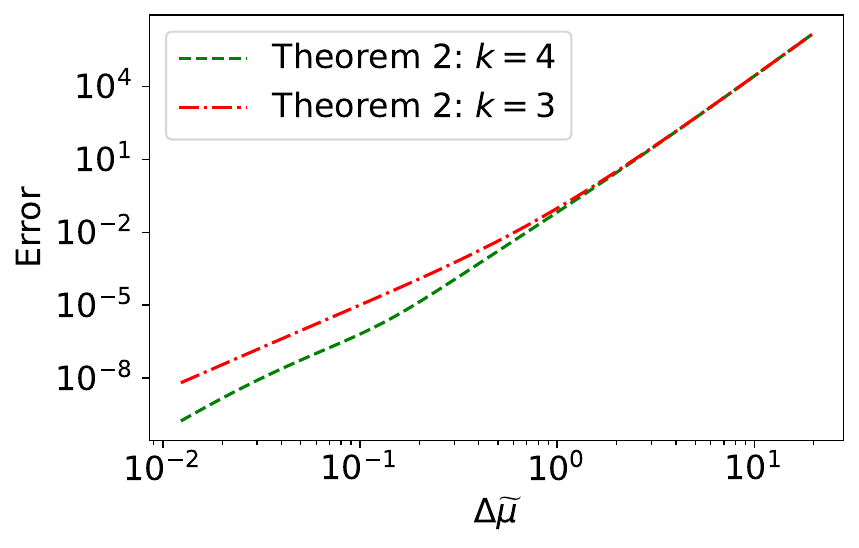}
\caption{Comparison of the computation of $\langle G\rangle_{\Delta T}$ (left) using the expansion \req{eq:T_faster_expansion_final} for $\gamma=1/2$, $d=1$, and $G(p)=E$, with the result obtained via numerical integration of \req{eq:T_faster_orig_integral}. The right plot shows the difference between the numerical result and the expansion \req{eq:T_faster_expansion_final} corresponding to $k=3$ and $k=4$. The right plot shows the difference between the numerical result and the two expansions. On a log-log plot the slopes of the error can be measured to be approximately $4$ (for $k=4$) and $7/2$ (for $k=3$), correctly matching the order of the expansion error in $\Delta\widetilde{\mu}$ as stated in \req{eq:T_faster_expansion_final}.}\label{fig:T_faster_expansion_comparison}
\end{figure}

 \subsection{ Sommerfeld Expansion in the Relativistic, Dense Regime $\mu\gg T\gg m$}\label{Section:HighTempSommerfeld} 
Interestingly, there is also a high temperature regime ($\mu\gg T\gg m$) where the traditional Sommerfeld expansion provides a good approximation to thermal averages. 
\begin{theorem}[High-Temperature Sommerfeld Expansion]\label{thm:high_T_Sommerfeld_main}
Let $k\in\mathbb{Z}^+$ be odd and suppose that $H(E)$ is continuous on $[m,\infty)$, is smooth on a neighborhood of $\infty$, and
\begin{align}\label{eq:H_deriv_asymp}
 H^{(k)}(E)=a_kE^{q_k}+\mathcal{O}(E^{q_k-1}) 
\end{align}
as $E\to \infty$ for some $a_k\neq 0$, $q_k\in\mathbb{R}$. Let $\mu=m[1+B(T/m)^{1+\gamma}]$, $\gamma,B>0$. Then
\begin{align}
 &\int_m^\infty dE\, H(E)f_{FD}(E)\notag\\
=&\int_m^\mu dE\, H(E)+\sum_{n=1,n\text{ odd}}^{k-2} 2(1-2^{-n})\zeta(n+1)T^{n+1}H^{(n)}(\mu)+\mathcal{O}(\mu^{q_k}T^{k+1})\label{eq:Sommerfeld_high_temp_final_main}
\end{align}
as $T\to\infty$.
\end{theorem}

To prove \req{eq:Sommerfeld_high_temp_final_main}, parameterize $\mu$ as a function of $T$ as follows
\begin{align}\label{eq:mu_infinity_faster}
\mu=m[1+B(T/m)^{1+\gamma}] \,,\,\,\gamma,B>0\,. 
\end{align}
and assume that the density of states $H(E)$ has the properties stated in Theorem \ref{thm:high_T_Sommerfeld_main}.
By integrating \req{eq:H_deriv_asymp} and combining it with the assumption that $H$ is continuous we can conclude that $H$ is also polynomially bounded on $[m,\infty)$. We note that, by definition, the asymptotic behavior \req{eq:H_deriv_asymp} implies that there exists $E_0$ and $C$ such that $H$ is smooth on $[E_0,\infty)$ and $|H^{(k)}(E)-a_kE^{q_k}|\leq CE^{q_k-1}$ for all $E\in[E_0,\infty)$. For the remainder of this derivation we assume that $T$ is large enough to ensure $\mu\geq E_0$. 

Now consider the nonzero temperature component of the decomposition \req{Eq_form} and use \req{eq:xpos} and \req{eq:xneg} to write
\begin{align}
 &\int_m^\infty dE\, H(E)(f_{T\neq 0}+\widetilde{f}_{T\neq 0})\notag\\
 =&\int_\mu^\infty dE \,H(E)\frac{1}{1+e^{(E-\mu)/T}}-\int_m^\mu dE\, H(E)\frac{1}{1+e^{(\mu-E)/T}}\notag\\
=&T\int_0^\infty dz \,H(\mu+Tz)\frac{1}{1+e^{z}}-T\int^{(\mu-m)/T}_0 dz\, H(\mu-Tz)\frac{1}{1+e^{z}}\,,\label{eq:Sommerfeld_high_temp_z}
\end{align}
where we changed variables to $z=(E-\mu)/T$ in the first integral and $z=(\mu-E)/T$ in the second. Due to the exponentially decaying factors in the integrands in \req{eq:Sommerfeld_high_temp_z} along with polynomial boundedness of $H$, we can shrink the upper limits of integration to $(\mu-E_0)/T$, thereby ensuring that $\mu-Tz\geq E_0$ on the domain of integration, with the error incurred by this change being $\mathcal{O}(T^{-r})$ as $T\to \infty$ for any choice of $r>0$. A similar bound was derived in the traditional Sommerfeld case, see Section \ref{sec:traditional_Sommerfeld} for details, though here this error term also accounts for the dependence on $\mu$ in accordance with \req{eq:mu_infinity_faster}. As we will see, the $\mathcal{O}(T^{-r})$ error term will be negligible compared to the other asymptotic approximations made below. In light of this we can write
\begin{align}
 &\int_m^\infty dE\, H(E)(f_{T\neq 0}+\widetilde{f}_{T\neq 0})\notag\\
=&T\int_0^{\frac{\mu-E_0}{T}} dz \,\frac{H(\mu+Tz)- H(\mu-Tz)}{1+e^{z}}+\mathcal{O}(T^{-r})\,. \label{eq:Sommerfeld_high_temp_z2}
\end{align}
As $\mu\pm Tz\geq E_0$ for all $z$ in the domain of integration, we can compute a Taylor expansion with remainder up to order $k$ on this domain as follows:
\begin{align}\label{eq:H_diff_Taylor}
 &H(\mu+Tz)- H(\mu-Tz)=\sum_{n=1,n\text{ odd}}^{k-2} 2H^{(n)}(\mu)\frac{(Tz)^n}{n!}\\
 &\qquad\qquad\qquad+(Tz)^k \int_0^1 ds\,\frac{(1-s)^{k-1}}{(k-1)!} \left[H^{(k)}(\mu+sTz)+H^{(k)}(\mu-sTz)\right]\,. \notag
\end{align}
We emphasize that the reason for restricting the domain of integration in \req{eq:Sommerfeld_high_temp_z2} (with the difference being accounted for in the error term) is that we are only assuming $H$ is $C^k$ on a neighborhood of $\infty$, and so the Taylor expansion argument is only valid on the restricted domain.

Substituting \req{eq:H_diff_Taylor} this into \req{eq:Sommerfeld_high_temp_z2} we can compute
\begin{align}
 &\int_m^\infty dE\, H(E)(f_{T\neq 0}+\widetilde{f}_{T\neq 0})\notag\\
 =&\sum_{n=1,n\text{ odd}}^{k-2} \frac{2T^{n+1}}{n!}H^{(n)}(\mu)\int_0^{\frac{\mu-E_0}{T}} dz \,\frac{z^n}{1+e^{z}}+R_k
+\mathcal{O}(T^{-r})\notag\\
=&\sum_{n=1,n\text{ odd}}^{k-2} 2(1-2^{-n})\zeta(n+1)T^{n+1}H^{(n)}(\mu)
+R_k
+\mathcal{O}(T^{-r})\,,\label{eq:Sommerfeld_high_temp_z3}
\end{align}
where to obtain the last line we again made use of the freedom to change the limits of integration while only incurring $\mathcal{O}(T^{-r})$ error and then used the integral formula \req{eq:FD_power_integrals}. The integrated remainder term is given by
\begin{align}
 R_k\equiv T^{k+1}\int_0^{\frac{\mu-E_0}{T}} dz \,\frac{z^k}{1+e^{z}}
 \int_0^1 ds\, \frac{(1-s)^{k-1}}{(k-1)!} \left[H^{(k)}(\mu+sTz)+H^{(k)}(\mu-sTz)\right]\,.
\end{align}
\req{eq:H_deriv_asymp} implies that $|H^{(k)}(E)|\leq CE^{q_k}$ on $[E_0,\infty)$ for some $C>0$. Using this we can bound the integral of the remainder as follows:
\begin{align}
 |R_k|\leq &CT^{k+1}\int_0^{\frac{\mu-E_0}{T}} dz \,\frac{z^k}{1+e^{z}}
 \int_0^1 ds\, \frac{(1-s)^{k-1}}{(k-1)!} \left(|\mu+sTz|^{q_z}+|\mu-sTz|^{q_k}\right)\notag\\
 \leq&2^{q_k+1}CT^{k+1}\int_0^{\infty} dz \,\frac{z^k}{1+e^{z}}\notag
 \int_0^1 ds\, \frac{(1-s)^{k-1}}{(k-1)!} \left[\mu^{q_k}+(sTz)^{q_k}\right]\notag\\
 =&\mathcal{O}(\mu^{q_k}T^{k+1})\label{eq:Sommerfeld_high_T_remainder_bound}
\end{align}
as $T\to\infty$. Note that due to the relation \req{eq:mu_infinity_faster} between $\mu$ and $T$ we can equivalently write the error term as $\mathcal{O}(T^{(1+\gamma)q_k+k+1})$ and so the $\mathcal{O}(T^{-r})$ error term is negligible in comparison. Combining the bound \req{eq:Sommerfeld_high_T_remainder_bound} with \req{eq:Sommerfeld_high_temp_z3} we obtain 
\begin{align}
 &\int_m^\infty dE\, H(E)f_{FD}(E)\notag\\
=&\int_m^\mu dE\, H(E)+\sum_{n=1,n\text{ odd}}^{k-2} 2(1-2^{-n})\zeta(n+1)T^{n+1}H^{(n)}(\mu)+\mathcal{O}(\mu^{q_k}T^{k+1})\label{eq:Sommerfeld_high_temp_final}
\end{align}
as $T\to\infty$. This completes the proof of \rTh{thm:high_T_Sommerfeld_main}.

Unless $q_k$ is sufficiently negative, the absolute error in \req{eq:Sommerfeld_high_temp_final_main} grows as $T\to\infty$, however the relative error decays as we now show. First assume $q_k>0$ and integrate \req{eq:H_deriv_asymp} twice at the Fermi energy $E=\mu$ to get
\begin{align}\label{eq:integrate_H_asymp}
 H^{(k-2)}(\mu)=\frac{a_k}{(q_k+1)(q_k+2)}\mu^{q_k+2}+\mathcal{O}(\mu^{q_k+1})
\end{align}
as $\mu\to\infty$. Therefore, the ratio of the error to the last included term in the expansion behaves as
\begin{align}\label{eq:Sommerfeld_high_T_rel_err}
&\frac{\mathcal{O}(\mu^{q_k}T^{k+1})}{ T^{k-1} H^{(k-2)}(\mu)}=\frac{T^2}{\mu^2}\frac{\mathcal{O}(1)}{ \frac{a_k}{(q_k+1)(q_k+2)}+\mathcal{O}(\mu^{-1})}= \mathcal{O}(T^{-2\gamma})\,,
\end{align}
which approaches $0$ as $T\to \infty$. To obtain \req{eq:Sommerfeld_high_T_rel_err} we used \req{eq:mu_infinity_faster} and also that $a_k\neq 0$. Continuing to integrate \req{eq:integrate_H_asymp}, one sees that earlier terms in the sum dominate the error term to a greater degree. Therefore the expansion \req{eq:Sommerfeld_high_temp_final} properly captures leading and sub-leading behavior, with a subdominant error term and a decaying relative error when $q_k>0$. When $q_k\leq 0$, integrating \req{eq:H_deriv_asymp} becomes more subtle due to the need to handle potential logarithm and constant terms that arise. However, if instead we make the additional assumptions 
\begin{align}
H^{(k-j)}(E)=a_{k-j} E^{q_k+j} +\mathcal{O}(E^{q_k+j-1})
\end{align}
for $j=2,...,k-1$ with $a_{k-j}\neq 0$ then a similar computation to \req{eq:Sommerfeld_high_T_rel_err} shows
\begin{align}
 \frac{\mathcal{O}(\mu^{q_k}T^{k+1})}{T^{n+1}H^{(n)}(\mu)}= \mathcal{O}(T^{-(k-n)\gamma})
\end{align}
for all $n=1,...,k-2$ and therefore the error term in \req{eq:Sommerfeld_high_temp_final} is again seen to be dominated by all other terms in the summation as $T\to\infty$. 

\begin{figure} 
\centering
\includegraphics[width=0.625\textwidth]{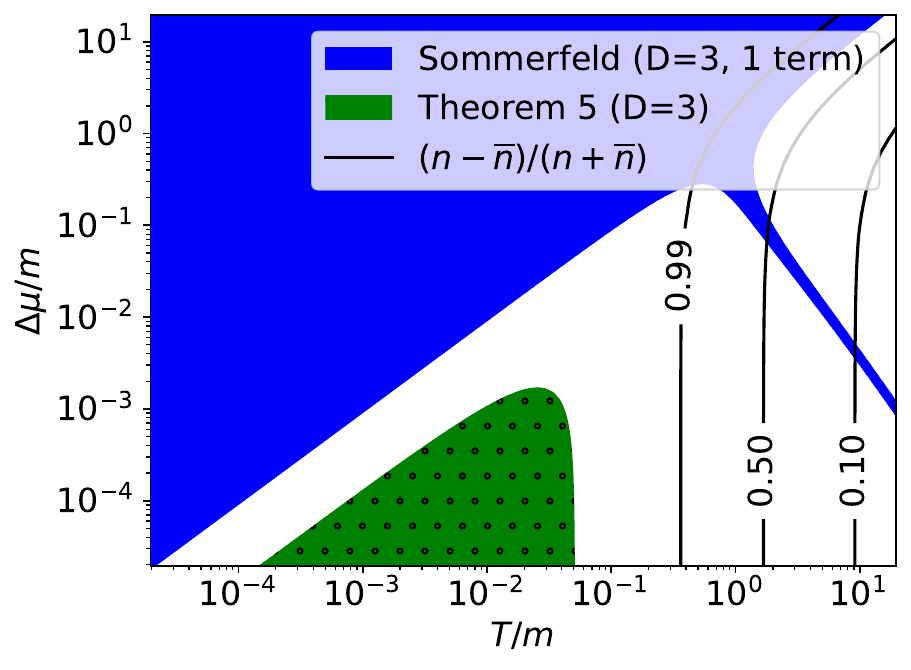}\\
\includegraphics[width=0.625\textwidth]{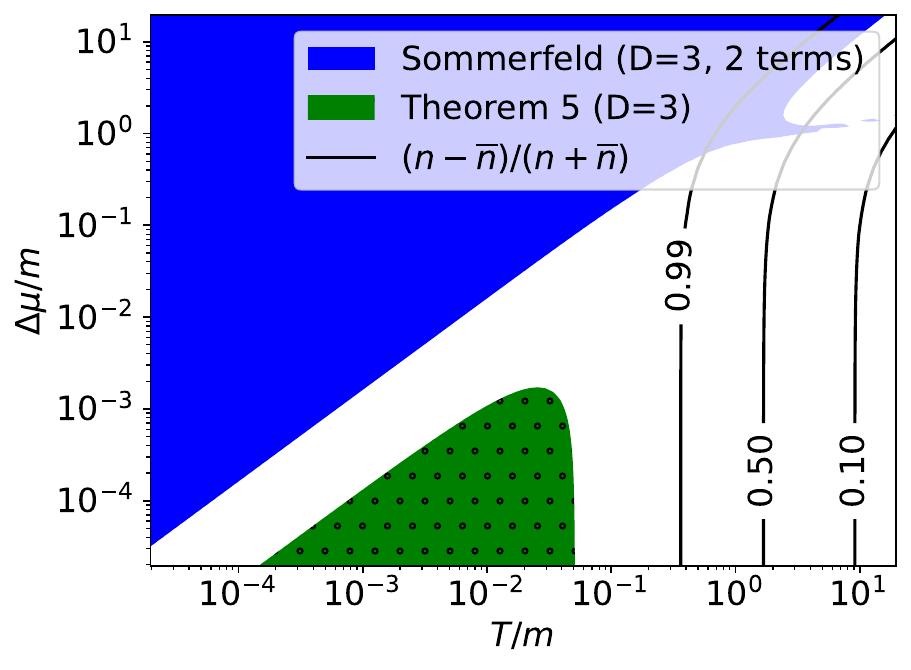}
\caption{Comparison of domains of validity for $D=3$ in $(T/m,\Delta\mu/m)$-space ($\Delta\mu\equiv\mu-m$), including relativistic parameter range, considering in top frame one asymptotic and in bottom frame two asymptotic terms; we show the Sommerfeld expansion (solid blue), and the expansion from \rTh{thm:mu_zero_faster} (dotted green). Validity is defined as the computation of $\langle G\rangle_T$ having relative error less that $10\%$ for the choice $G(p)=1$. See text for further details.}\label{fig:Thm3_vs_Sommerfeld_regions_terms_comp_part_antipart}
\end{figure}

In \rf{fig:Thm3_vs_Sommerfeld_regions_terms_comp_part_antipart} we demonstrate numerically that the Sommerfeld expansion from \rTh{thm:high_T_Sommerfeld_main} applies to the high-temperature region, in addition to the well-known low temperature regime. In contrast, our new expansion, see below \rTh{thm:mu_zero_faster}, applies when $|\Delta\mu|\ll T\ll m$, a regime not covered by either the Sommerfeld expansion or Boltzmann approximation; for a comparison with the latter see \rf{fig:Thm3_vs_Sommerfeld_regions}. The domians of validity of both expansions predominantly lie in the particle dominated region where $n\gg \overline{n}$ ($n$ and $\overline{n}$ denote number of particles and antiparticles respectively); see the solid black contours. The pair dominated region ($\overline{n}\approx n$) on the right side of the domain requires other methods. We note that the domain of validity when using 2 terms (bottom frame \rf{fig:Thm3_vs_Sommerfeld_regions_terms_comp_part_antipart}) of the Sommerfeld expansion is reduced in many areas, as compared to the region for 1 term (top frame). This is typical of results such as \rTh{thm:high_T_Sommerfeld_main} that give asymptotic, but not convergent, expansions; adding additional terms does not necessarily increase the accuracy at fixed values of the parameters $\mu$ and $T$. The leg structure of the blue domain arises since the approximate and exact results cross, thus this reflects only an accidental exact agreement between the numerical and one-term asymptotic results.

\section{Non-Sommerfeld Few Particle Regime}\label{sec:beyondSommer}
\subsection{Asymptotic Expansion in the Regime $T\to 0$ with $\mu-m\ll T$}\label{sec:asympt_Delta_mu_order_T}
In this section we study thermal averages when $\mu-m\ll T\ll m$, a regime where we demonstrate a complete failure of the Sommerfeld formula. The techniques we use to study this regime are similar to those employed in Section \ref{sec:asymp_T_0_faster}, but this time they lead to a wholly distinct and novel asymptotic expansion. In \rf{fig:FD_avg_expansion_comparison_Sommerfeld} we preview this new result, which is stated in \rTh{thm:mu_zero_faster} below. In this figure we have parameterized $\mu-m=\mathcal{O}(T^{3/2})$ and so $\mu-m$ approaches zero faster than $T$. Note that here the Sommerfeld expansion \req{eq:traditional_Sommerfeld_final} (red dot-dashed lines) fails at low temperature while our \rTh{thm:mu_zero_faster} (green dashed lines) provides an accurate approximation. One can see from the figure that the traditional Sommerfeld expansion is in fact a high temperature approximation under this relation between $\mu$ and $T$; this corresponds to the behavior proven in \rTh{thm:high_T_Sommerfeld_main} above.

\begin{figure}[h] 
\centering
\includegraphics[width=0.49\textwidth]{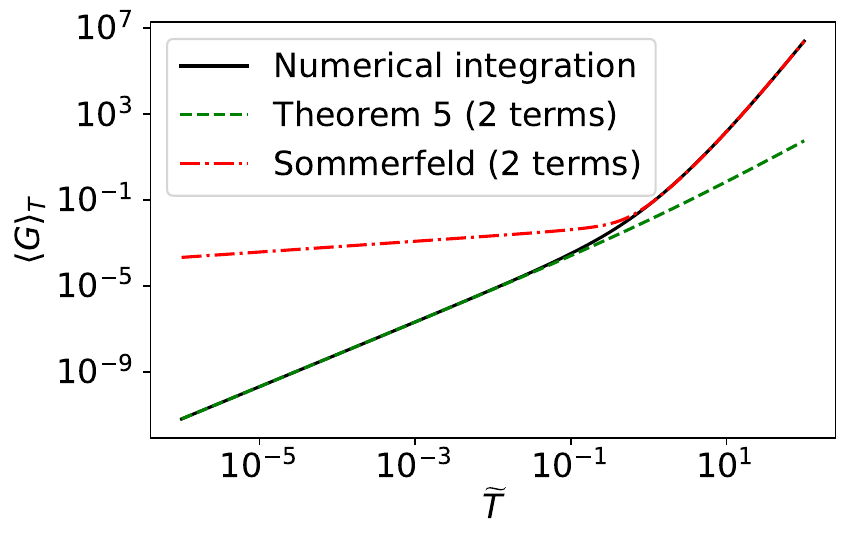}
\includegraphics[width=0.49\textwidth]{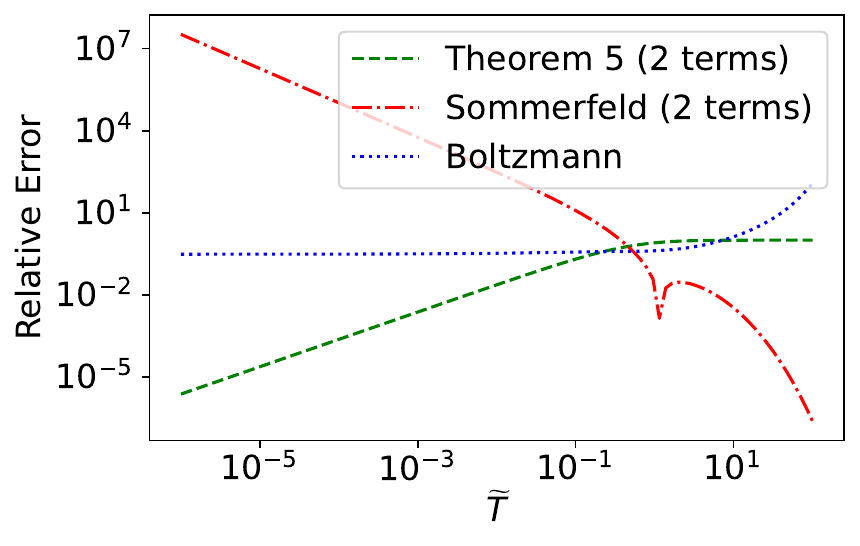}
\caption{Left: Comparison of asymptotic expansions of $\langle G\rangle_{T}$, $G(p)=1$, via the Sommerfeld expansion \req{eq:traditional_Sommerfeld_final} (red dot-dashed line) and the new expansion \req{eq:asymp_FD_int_b_decay} (green dashed line) where $\mu=m(1+\widetilde{T}^{3/2})$. Under this relationship between $\mu$ and $T$, \rTh{thm:high_T_Sommerfeld_main} and \rTh{thm:mu_zero_faster} respectively prove that \req{eq:Sommerfeld_high_temp_final_main} is valid at high temperature and \req{eq:asymp_FD_int_b_decay} is valid at low temperature; we see this reflected in the figure. Right: The relative error of these asymptotic expansions, also compared with that of the Boltzmann approximation. }\label{fig:FD_avg_expansion_comparison_Sommerfeld}
\end{figure}

We now proceed with a detailed derivation of the new expansion. To make our assumptions explicit, here we consider a $D$-dimensional thermal average ($D\geq 2$) of a function $G(p)$, $p\in[0,\infty)$ that is $C^k$ and whose zeroth through $k$'th derivatives are polynomially bounded, and will assume that $\mu=m+bT$, $b\in\mathbb{R}$. To begin the derivation, first change variables in \req{eq:avg_G_T} to $z=(E-m)/m$ so that $p=m\sqrt{z}\sqrt{z+2}$:
\begin{align}
&\int_0^\infty dp\, p^{D-1} G(p) f_{FD}(p)\notag\\
=&m^{D}\int_0^\infty dz\, (1+z) z^{(D-2)/2} (z+2)^{(D-2)/2} G\left(m\sqrt{z}\sqrt{z+2}\right) \frac{1}{1+e^{z/\widetilde{T}-b}}\,,\label{eq:fd_int_var_z}
\end{align}
where $\widetilde{T}\equiv T/m$. Note that the integrand can have a square root singularity at $z=0$ and so we cannot simply Taylor expand in $z$; this is a key difference from the derivation of the traditional Sommerfeld expansion. However we can employ a more general asymptotic expansion by again making use of the function $F(y,m)$, defined in \req{eq:F_y_m_def}, and its expansion \req{eq:F_y_m_Taylor}, to rewrite the integral as follows:
\begin{align}
 \req{eq:fd_int_var_z}=&m^{D}\int_0^\infty dz\, F(\sqrt{z},m)\frac{z^{(D-2)/2}}{1+e^{z/\widetilde{T}-b}}\\
 =&\sum_{n=0}^{k-1} a_n(m)m^{D}\int_0^\infty dz\, \frac{z^{(n+D-2)/2}}{1+e^{z/\widetilde{T}-b}}+m^{D}\int_0^\infty dz\, R_k(\sqrt{z},m)\frac{z^{(D-2)/2}}{1+e^{z/\widetilde{T}-b}}\,.\notag
\end{align}

When $b\geq 0$ the integral of the remainder term can be bounded as follows
\begin{align}
 &\left|m^{D}\int_0^\infty dz\, R_k(\sqrt{z},m)\frac{z^{(D-2)/2}}{1+e^{z/\widetilde{T}-b}}\right|\\
 \leq&m^{D}\int_0^\infty dz\, [\alpha_k(m)+\beta_k(m)z^{q_k/2}]\frac{z^{(k+D-2)/2}}{1+e^{z/\widetilde{T}-b}}\notag\\
 \leq &m^{D}\int_{b\widetilde{T}}^\infty dz\, [\alpha_k(m)+\beta_k(m)z^{q_k/2}]z^{(k+D-2)/2}e^{-z/\widetilde{T}+b}\notag\\
 &+m^{D}\int_0^{{b\widetilde{T}}} dz\, [\alpha_k(m)+\beta_k(m)z^{q_k/2}]z^{(k+D-2)/2}\notag\\
 =&m^{D} \alpha_k(m)\widetilde{T}^{(k+D)/2}\int_{0}^\infty du\,(u+b)^{(k+D-2)/2}e^{-u}\notag\\
 &+m^D\beta_k(m)\widetilde{T}^{(k+D+q_k)/2}\int_{0}^\infty du\,(u+b)^{(k+D-2+q_k)/2}e^{-u}\notag\\
 &+\alpha_k(m)m^{D}\frac{2}{k+D}(b\widetilde{T})^{(k+D)/2}+\beta_k(m)m^{D}\frac{2}{k+D+q_k}(b\widetilde{T})^{(k+D+q_k)/2}\notag\\
 =&\mathcal{O}(\widetilde{T}^{(k+D)/2})\label{eq:combined_exp_remainder}
\end{align}
as $\widetilde{T}\to 0$. When $b<0$ the integral of the remainder can be bounded more simply as follows
\begin{align}
 &\left|m^{D}\int_0^\infty dz\, R_k(\sqrt{z},m)\frac{z^{(D-2)/2}}{1+e^{z/\widetilde{T}-b}}\right|\\
 \leq&m^{D}\int_0^\infty dz\, (\alpha_k(m)+\beta_k(m)z^{q_k/2}) 
z^{(k+D-2)/2}e^{-z/\widetilde{T}+b}\notag\\
 \leq&\alpha_k(m)m^{D}e^b\widetilde{T}^{(k+D)/2}\int_0^\infty du\, u^{(k+D-2)/2}e^{-u}\notag\\
 &+\beta_k(m)m^{D}e^b\widetilde{T}^{(q_k+k+D)/2}\int_0^\infty du\,u^{(q_k+k+D-2)/2}e^{-u} 
=\mathcal{O}(\widetilde{T}^{(k+D)/2}) \label{eq:combined_exp_remainder2}
\end{align}
as $\widetilde{T}\to 0$.
\begin{remark}\label{remark:combined_remainder_uniform_in_b}
We emphasize that the implied constant in the bounds \req{eq:combined_exp_remainder} and \req{eq:combined_exp_remainder2} depends on $m$ and $b$, though if one restricts $b$ to a set that is bounded above then the constants can be chosen independent of $b$. 
\end{remark}

Changing variables to $x=z/\widetilde{T}$, recalling the integral formula for the polylogarithm function $\mathrm{Li}_s(y)$, see 3.411.3 in \cite{Gradshteyn:1943cpj},
\begin{align}\label{eq:h_decomp_eval}
 \int_0^\infty dx \, \frac{x^{n/2}}{1+e^{x-b}} =-\Gamma(1+n/2)\mathrm{Li}_{1+n/2}(-e^b)\,,
\end{align}
and using the remainder bounds \req{eq:combined_exp_remainder} and \req{eq:combined_exp_remainder2} we obtain the following:
\begin{theorem}\label{thm:asymp_FD_int_Delta_mu_small}
Let $D\geq 2$, $k\in\mathbb{Z}^+$, and $G(p)$ be a $C^k$ function on $[0,\infty)$ whose zeroth through $k$'th derivatives are polynomially bounded. Let $\mu=m+bT$ for $b\in\mathbb{R}$. Then
\begin{align}\label{eq:asymp_FD_int}
&\int_0^\infty dp\, p^{D-1} G(p) f_{FD}(p)\notag\\
 =-&\sum_{n=0}^{k-1} a_n(m)m^{D}\widetilde{T}^{(n+D)/2}\Gamma[(n+D)/2]\mathrm{Li}_{(n+D)/2}(-e^b)+\mathcal{O}\left(\widetilde{T}^{(k+D)/2}\right)\,,
\end{align}
as $\widetilde{T}\to 0$, where $\widetilde{T}=T/m$, $a_n$ is defined in \req{eq:an_def} (explicit formulas for the first few $a_n$'s can be found in \req{eq:first_few_a_ns}), and the implied constant in the error term depends on $m$ but is uniform in $b$ when restricted to a fixed set that is bounded above. 
\end{theorem}

\subsection{Asymptotic Expansion in the Regime $T\ll m$, $\mu=m$}\label{sec:b_0}
When $b=0$ (i.e., $\mu=m$) one can use the integral formula \req{eq:FD_power_integrals} to further simplify the expansion \req{eq:asymp_FD_int}. In dimension $D>2$ we obtain
\begin{align}\label{eq:asymp_FD_int_b_0}
&\int_0^\infty dp\, p^{D-1} G(p) f_{FD}(p)\\
 =&\sum_{n=0}^{k-1} a_n(m)m^{D}\widetilde{T}^{(n+D)/2}\left(1-2^{-(n+D-2)/2}\right)\Gamma[(n+D)/2]\zeta[(n+D)/2]\notag\\
 &+\mathcal{O}\left(\widetilde{T}^{(k+D)/2}\right)\notag
\end{align}
as $\widetilde{T}\to 0$, while for $D=2$ we have
\begin{align}\label{eq:asymp_FD_int_b_D2}
&\int_0^\infty dp\, p G(p) f_{FD}(p)\\
 =& a_0(m)m^2 \ln(2)\widetilde{T}+ 
 \sum_{n=1}^{k-1} a_n(m)m^{2}\widetilde{T}^{(n+2)/2}\left(1-2^{-n/2}\right)\Gamma[(n+2)/2]\zeta[(n+D)/2]\notag\\
 &+\mathcal{O}\left(\widetilde{T}^{(k+2)/2}\right)\notag
\end{align}
as $\widetilde{T}\to 0$. In particular, note that when $D=3$ and $G(0)\neq 0$ the leading order term scales with $\widetilde{T}^{3/2}$. This is in contrast to the Sommerfeld expansion \req{eq:traditional_Sommerfeld_final}, which applies to the case where $\mu-m$ is bounded away from zero and whose leading order term scales with $\widetilde{T}^2$.

\subsection{Asymptotic Expansion in the Regime $|\mu-m|\ll T\ll m$}\label{sec:decaying_b}
One can also use the expansion \req{eq:asymp_FD_int} to study the regime where $\mu$ approaches $m$ faster than $T$ approaches zero. Mathematically we parameterize this as 
\begin{align}\label{eq:mu_decay_faster}
\mu=m(1+B\widetilde{T}^{1+\gamma})\,,\,\,\, B\in\mathbb{R}\,,\,\, \gamma>0\,. 
\end{align}
 To proceed, first Taylor expand \req{eq:h_decomp_eval} to first order in $b$ by differentiating under the integral to obtain
\begin{align}\label{eq:Li_expansion}
 &-\Gamma(1+n/2)\mathrm{Li}_{1+n/2}(-e^b) =\int_0^\infty du \, \frac{u^{n/2}}{1+e^{u-b}} \\
 =&\int_0^\infty du\,\frac{u^{n/2}}{e^u+1}+b\int_0^\infty du\,\frac{u^{n/2}}{(e^{u/2}+e^{-u/2})^2} +\mathcal{O}(b^2)\notag\\
 =&(1 - 2^{-n/2}) \Gamma(1 + n/2) \zeta(1 + n/2)+ b\int_0^\infty du\,\frac{u^{n/2}}{(e^{u/2}+e^{-u/2})^2} +\mathcal{O}(b^2)\,.\label{eq:Li_expansion_final}
\end{align}
See \cite{dingle1957fermi} and \cite{10.1063/1.1350634} for in-depth discussion of the computation of \req{eq:Li_expansion}. For $n=0$ the leading order term in \req{eq:Li_expansion_final} has the indeterminate form $0\cdot \infty$ (this case will only be relevant to us when $D=2)$; there one should instead use
\begin{align}
\int_0^\infty du \frac{1}{1+e^{u-b}}=\ln(1+e^b)= \ln(2)+\frac{b}{2}+\mathcal{O}(b^2)\,.
\end{align}
{\xred 
The integral in \req{eq:Li_expansion_final} can be evaluated via integration by parts:
\begin{align}\label{eq:Li_int_order_1}
 \int_0^\infty du\,\frac{u^{n/2}}{(e^{u/2}+e^{-u/2})^2}=&-u^{n/2}/(e^u+1)|_0^\infty+ \frac{n}{2}\int_0^\infty du\, \frac{u^{n/2-1}}{e^{u}+1}\\
 =&\frac{n}{2}\begin{cases}
 (1-2^{1-n/2})\Gamma(n/2)\zeta(n/2)& n>0,\,n\neq 2\\
 \log(2) &n=2
 \end{cases}\,.\notag
 \end{align}}
Now substitute \req{eq:Li_expansion} into \req{eq:asymp_FD_int} and let $b=B\widetilde{T}^\gamma$. We note that the remainder term in \req{eq:asymp_FD_int} is still $\mathcal{O}(\widetilde{T}^{1+(k+D-2)/2})$ in this case because $b$ remains bounded above (in fact, it remains bounded) as $\widetilde{T}\to 0$; see Remark \ref{remark:combined_remainder_uniform_in_b}. The expansion \req{eq:asymp_FD_int} with $k=2$ then simplifies to give the following result.
\begin{theorem}\label{thm:mu_zero_faster}
Let $G(p)$ be a $C^2$ function on $[0,\infty)$ whose zeroth through $2$nd derivatives are polynomially bounded. Let $\mu=m(1+B\widetilde{T}^{1+\gamma})$ where $B\in\mathbb{R}$, $\gamma>0$, $\widetilde{T}=T/m$. 
\begin{enumerate}
 \item {\xred For $D>2$ and with $a_n$ defined by \req{eq:an_def} we have
\begin{align}
&\int_0^\infty dp\, p^{D-1} G(p) f_{FD}(p)\label{eq:FD_b_decay_orig_integral}\\
 =&a_0(m)m^{D}(1 - 2^{-(D-2)/2}) \Gamma( D/2) \zeta(D/2)\widetilde{T}^{D/2}\notag\\
 &+ a_0(m)m^{D}B\widetilde{T}^{D/2+\gamma}\frac{D-2}{2}\begin{cases}
 (1-2^{2-D/2})\Gamma[(D-2)/2]\zeta[(D-2)/2]& D\neq 4\\
 \log(2) &D=4
 \end{cases}\notag\\
 &+a_1(m)m^{D}(1 - 2^{-(D-1)/2}) \Gamma[(D+1)/2] \zeta[(D+1)/2]\widetilde{T}^{(D+1)/2}\notag\\
 &+\mathcal{O}\left(\widetilde{T}^{(2+D)/2}\right)+\mathcal{O}(\widetilde{T}^{D/2+2\gamma})\,.\label{eq:asymp_FD_int_b_decay}
\end{align}
}
\item For $D=2$ and with $a_n$ defined by \req{eq:an_def} we have
\begin{align}
&\int_0^\infty dp\, p G(p) f_{FD}(p) 
 =  a_0(m)m^{2}\ln(2)\widetilde{T}+\frac{1}{2}a_0(m)m^{2}B\widetilde{T}^{1+\gamma}\notag\\&+a_1(m)m^{2}(1 - 2^{-1/2}) \frac{\pi^{1/2}}{2} \zeta(3/2)\widetilde{T}^{3/2} +\mathcal{O}\left(\widetilde{T}^{1+2\gamma}\right)+\mathcal{O}\left(\widetilde{T}^{2}\right) \,\,\,\,\text{as}\,\, \,\widetilde{T}\to 0\,.\label{eq:asymp_FD_int_b_decay_D2}
\end{align}
\end{enumerate}
\end{theorem}
\begin{remark}
 When referring to the number of terms in the expansions in \rTh{thm:mu_zero_faster} and related results, e.g., in \rf{fig:FD_avg_expansion_comparison_Sommerfeld}, we mean the index $k$ that was used in \req{eq:asymp_FD_int} to obtain the expansion.
\end{remark}
{\xred\begin{remark}
Even when the physical dimension $D$ is  fixed, the general result \req{eq:FD_b_decay_orig_integral} for arbitrary $D$ can be used to simplify calculations in cases where $G$ has the form $G(p)=p^\kappa \widetilde G(p)$, $D+\kappa>2$, by replacing $D$ with $D+\kappa$ and $G$ with $\widetilde G$ in \req{eq:FD_b_decay_orig_integral} and \req{eq:an_def}. For instance, to compute the pressure in $D=3$ dimensions one can let $\kappa=2$ and $\widetilde{G}(p)= 1 / (3 E) $.
\end{remark}}

The dominant error term in \req{eq:asymp_FD_int_b_decay} and \req{eq:asymp_FD_int_b_decay_D2} depends on whether $\gamma<1/2$ or $\gamma\geq 1/2$. To obtain higher order terms in these expansions one simply needs to continue the expansion \req{eq:Li_expansion} to the appropriate higher order before substituting it into \req{eq:asymp_FD_int}; however, we again emphasize that these results are asymptotic expansions and so going to higher order does not necessarily increase the accuracy at a given $T$. When applying \rTh{thm:mu_zero_faster} to given values of $\mu$ and $T$, i.e., without reference to any parameterization of the form \req{eq:mu_decay_faster}, we fix $B=1$ if $\mu>m$ and $B=-1$ if $\mu<m$ and then choose the unique $\gamma$ that makes the equality \req{eq:mu_decay_faster} hold (the case $\mu=m$ was already considered in Section \ref{sec:b_0}). 

See \rf{fig:FD_avg_expansion_comparison_Sommerfeld} above for a comparison between the expansion \req{eq:asymp_FD_int_b_decay} (green dashed line) and numerical integration of \req{eq:FD_b_decay_orig_integral} (solid black line). We also compare with the Sommerfeld expansion \req{eq:traditional_Sommerfeld_final} (red dot-dashed line). Note that the Sommerfeld expansion completely fails in the low temperature regime when $\mu$ and $T$ are related by \req{eq:mu_decay_faster} with $0<\mu-m\ll T$. Under the parameterization \req{eq:mu_decay_faster}, the traditional Sommerfeld expansion is actually seen to be valid at high temperature; we proved this to be the case in \rTh{thm:high_T_Sommerfeld_main} above. 

In \rf{fig:Thm3_vs_Sommerfeld_regions_terms_comp_deBroglie} we again present a comparison between the domain of applicability of the Sommerfeld expansion and \rTh{thm:mu_zero_faster}. Here we show the number of particles per de Broglie volume in the solid black contours, from which we can see that our new expansion applies in a moderate density regime where quantum effects are non-negligible but that is distinct from the high-density regime where the Sommerfeld expansion applies. The region of validity of \rTh{thm:mu_zero_faster} is a bit larger in dimension $D=2$ (bottom frame) than in $D=3$ (top frame).

\begin{figure} 
\centering
\includegraphics[width=0.625\textwidth]{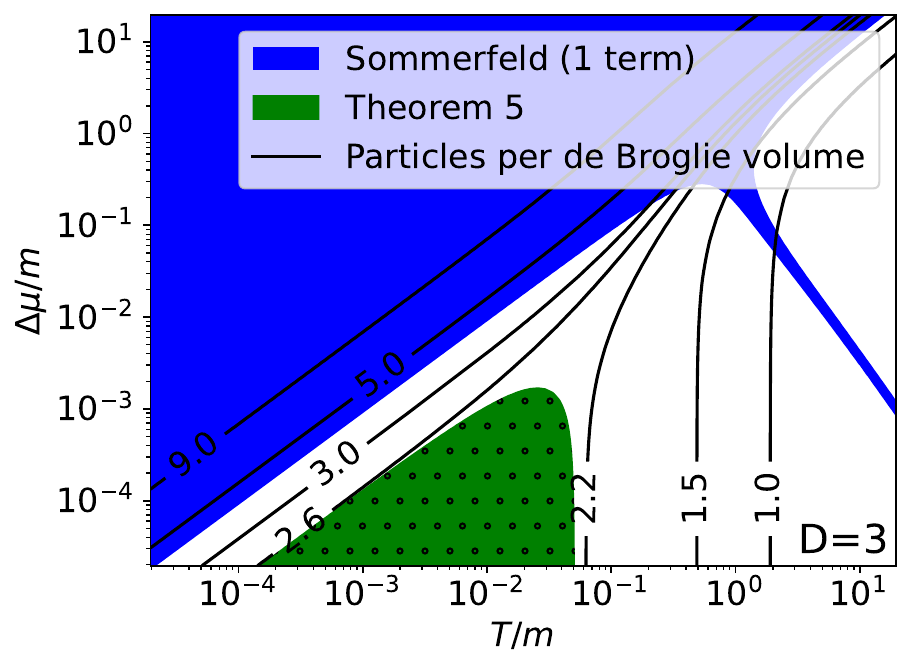}\\
\includegraphics[width=0.625\textwidth]{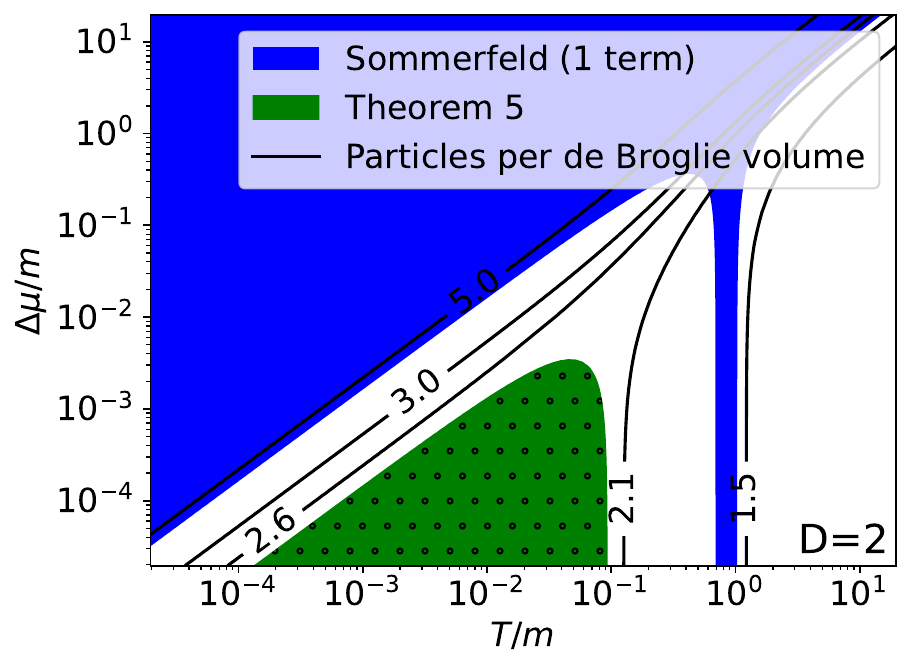}
\caption{Comparison of two and three dimensional cases of domains of validity in $(T/m,\Delta\mu/m)$-surface ($\Delta\mu\equiv\mu-m$), including relativistic parameter range, of the Sommerfeld expansion (solid blue) and the expansion from \rTh{thm:mu_zero_faster} (dotted green), where validity is defined as the computation of $\langle G\rangle_T$ having relative error less that $10\%$ for the choice $G(p)=1$. }\label{fig:Thm3_vs_Sommerfeld_regions_terms_comp_deBroglie}
\end{figure}

\section{Discussion of Results}
\label{sec:final}
We resolved a century long oversight, the understanding of the dilute Fermi-Dirac (FD) quantum domain in which the traditional Sommerfeld expansion fails. We recognize the particle number per de Broglie volume as a critical physical quantity which separates the different FD regimes, i.e., the regions in the $(T/m,(\mu-m)/m)$-plane where different asymptotic expansions apply. We obtain a novel asymptotic expansion that is valid when the particle number per de Broglie volume is moderate (2 to 3 particles), as compared to the low density semiclassical Boltzmann limit (1 particle or less) and the traditional highly degenerate Sommerfeld limit (5 or more particles); see \rf{fig:Thm3_vs_Sommerfeld_regions}. This intermediate density regime was unrecognized in the literature, where often one finds a mere acknowledgement that ``the Sommerfeld expansion fails''. In addition, we add to the understanding of the Sommerfeld expansion in the high temperature regime and when $\mu$ is close to $m$.

 Our main results in regard to different regimes are as follows, with detailed assumptions found in the referenced theorems, and figure illustrations offered for the nonrelativistic regime in \rf{fig:Thm3_vs_Sommerfeld_regions}, and extension to relativistic regime along with comparisons in \rf{fig:Thm3_vs_Sommerfeld_regions_terms_comp_part_antipart} and \rf{fig:Thm3_vs_Sommerfeld_regions_terms_comp_deBroglie}:
\begin{enumerate}
\item 
{\bf Nonrelativistic degenerate Fermi gas:} Regime $T\ll\mu-m\ll m$ considered for dimension $D\geq 2$, with $\Delta\widetilde{\mu}=(\mu-m)/m\to 0^+$, while $T$ is constrained by $T=d\Delta\widetilde{\mu}^{1+\gamma}m$, $d,\gamma>0$ (\rTh{thms:T_decay_faster}),
\begin{align}
 &\int_0^\infty dp\, p^{D-1} G(p)(f_{T\neq 0}+\widetilde{f}_{T\neq 0})=\sum_{n=0}^{k-1} da_n(m)m^{D}\Delta\widetilde{\mu}^{(n+D)/2+\gamma}\\
 &\times\sum_{j=1,\mathrm{odd}}^{\lceil(k-n)/(2\gamma)\rceil-1}
 2d^j\Delta\widetilde{\mu}^{j\gamma}\left(\prod_{\ell=0}^{j-1}[(n+D-2)/2-\ell] \right)(1-2^{-j})\zeta(j+1) \notag\\
 &+\mathcal{O}\left(\Delta\widetilde{\mu}^{(k+D)/2+\gamma}\right)\;,\notag\\
&a_n(m)=\frac{1}{n!}\partial_y^n|_{y=0}\left[ (y^2+1)( y^2+2)^{(D-2)/2} G\left(my\sqrt{y^2+2}\right)\right]\,.
\end{align}
Here $f_{T\neq 0}+\widetilde{f}_{T\neq 0}$ denotes the finite temperature component of the FD distribution as defined in \req{Eq_form}.
\item 
{\bf Relativistic, dense Fermi gas:} Regime $\mu\gg T\gg m$ with $T\to\infty$, while $\mu$ is constrained by $\mu=m[1+B(T/m)^{1+\gamma}]$, $\gamma,B>0$ (\rTh{thm:high_T_Sommerfeld_main}),
\begin{align}
 &\int_m^\infty dE\, H(E)f_{FD}(E)=\int_m^\mu dE\, H(E)\\
&+\sum_{n=1,n\text{ odd}}^{k-2} 2(1-2^{-n})\zeta(n+1)T^{n+1}H^{(n)}(\mu)+\mathcal{O}(\mu^{q_k}T^{k+1})\;.\notag
\end{align} 
\item 
{\bf Dilute quantum Fermi gas $D=3$:} Regime $|\mu-m|\ll T\ll m$ with $\widetilde T=T/m\to 0$, while $\mu$ is constrained by $\mu=m[1+B\widetilde{T}^{1+\gamma}]$, $B\in\mathbb{R}$, $\gamma>0$ (\rTh{thm:mu_zero_faster}, $D=3$),
\begin{align}
&\int_0^\infty dp\, p^{2} G(p) f_{FD}(p) =m^{3}G(0)(1 - 2^{-1/2}) \left(\frac{\pi}{2}\right)^{1/2} \!\!\zeta(3/2)\widetilde{T}^{3/2}\label{eq:thm5_D3_conc}\\
 &- m^{3}G(0)({2}^{1/2}-1)\left(\frac{\pi}{2}\right)^{1/2}\!\!\zeta(1/2)B\widetilde{T}^{3/2+\gamma} +2m^{4}G^\prime(0) \frac{\pi^2}{12}\widetilde{T}^{2}\notag\\
 &+\mathcal{O}(\widetilde{T}^{3/2+2\gamma})+ \mathcal{O}\left(\widetilde{T}^{5/2}\right)\;. \notag
\end{align} 
\item 
{\bf Dilute quantum Fermi gas $D=2$:} Regime $|\mu-m|\ll T\ll m$ with $\widetilde{T}=T/m\to 0$, while $\mu$ is constrained by $\mu=m[1+B\widetilde{T}^{1+\gamma}]$, $B\in\mathbb{R}$, $\gamma>0$ (\rTh{thm:mu_zero_faster}, $D=2$),
\begin{align}
&\int_0^\infty dp\, p G(p) f_{FD}(p)=m^{2}G(0)\ln(2)\widetilde{T} \label{eq:thm5_D2_conc}\\
 &+\frac{1}{2}m^{2}G(0)B\widetilde{T}^{1+\gamma}+m^{3}G^\prime(0)(1 - 2^{-1/2}) \left(\frac{\pi}{2}\right)^{1/2} \zeta(3/2)\widetilde{T}^{3/2}\notag\\
 &+\mathcal{O}\left(\widetilde{T}^{1+2\gamma}\right)+\mathcal{O}\left(\widetilde{T}^{2}\right)\;.\notag
\end{align} 
\end{enumerate}
Further terms in \req{eq:thm5_D3_conc} and \req{eq:thm5_D2_conc} can be obtained by using the more general result in \rTh{thm:asymp_FD_int_Delta_mu_small} {\xred and results for other values of $D$ are given in \rTh{thm:mu_zero_faster}. }

We have also introduced as a supplemental tool a novel form of the FD distribution \req{Eq_form} that separates the Fermi gas into zero and finite temperature components, see \rf{Fermi_Component}. We find this decomposition convenient when deriving asymptotic expansions of finite $T$ effects, as seen in Section \ref{sec:Sommerfeld}.

We note that the analysis presented here does not incorporate a number of features that are important in specific physics applications, such as magnetism, spin, particle-antiparticle effects, and cosmological free-streaming distributions. We intend to consider these natural refinements of the methods presented here in future application-oriented work.
 


\backmatter

\bmhead{Acknowledgments}
We thank Gordon Baym and John W. Clark for their encouragement to pursue research into the novel form of the Fermi distribution which ultimately lead to this work.

\input novel-fermi-function.bbl

\end{document}